\documentclass[%
 reprint,
 amsmath,amssymb,
 aps
]{revtex4-2}

\usepackage{graphicx}
\usepackage{dcolumn}
\usepackage{bm}
\usepackage{amsmath}
\usepackage{xcolor}
\usepackage{hyperref}
\usepackage{soul}

\begin{document}

\newcommand{\rg}[1]{\textcolor{magenta}{#1}}
\newcommand{\rgc}[1]{\textcolor{magenta}{{[\bf #1]}}}
\newcommand{\trc}{translation-rotation coupling }
\newcommand{\hj}[1]{\textcolor{black}{#1}}

\preprint{APS/123-QED}

\title{\hj{Sedimentation Dynamics of Bodies with Two Planes of Symmetry}}

\author{Harshit Joshi}
\author{Rama Govindarajan}%
 \email{rama@icts.res.in}
\affiliation{%
 International Centre for Theoretical Sciences, Bengaluru, 560089 India
}%


\begin{abstract}

We show that bodies with two planes of symmetry can display a range of behaviours even without inertia. Any such body supports a conserved quantity in its dynamics, and is either a settler, a drifter or a flutterer, depending only on its shape. At large time, settlers and drifters respectively fall vertically and obliquely, while flutterers rotate forever while executing intricate patterns. The dynamics of flutterers decouples into a periodic and a Floquet part with different time scales, giving periodicity or quasi-periodicity. 
We design a set of bodies and use the boundary integral method to show that settlers, drifters and flutterers, all lie in this set. 

\vspace{0.5cm}
\noindent DOI: \href{https://journals.aps.org/prl/abstract/10.1103/PhysRevLett.134.014002}{10.1103/PhysRevLett.134.014002}

\end{abstract}

\maketitle

The complex and poetic dance of falling autumn leaves is a high Reynolds number phenomenon, which depends crucially on vortex shedding \cite{field1997chaotic, zhong2011experimental, auguste2013falling}, a feature absent in Stokes flow. One would therefore expect settling at zero Reynolds number to be uneventful, especially for a body shape of high symmetry. 
In fact, in steady Stokes flow, the linear and angular motion of bodies with three planes of symmetry, like ellipsoids, are completely decoupled, and without external torque can only display linear motion \cite{happel2012low,kim2013microhydrodynamics,witten2020review, guazzelli2011physical, graham2018microhydrodynamics}. 
But particles can have complex shapes with fewer symmetries, and
understanding their sedimentation is crucial in diverse contexts, from marine snow sequestering atmospheric carbon to ice crystals in clouds and industrial powder production, like milk powder.
Unlike ellipsoids,
bodies with fewer symmetries can couple translation and rotation, opening up richer possibilities, e.g., chiral sedimenting trajectories of chiral objects like helices and Kelvin's isotropic helicoid \cite{PhysRevFluids.3.124301,krapf2009chiral, collins2021lord}. 
The pioneering work of Brenner \cite{brenner1964stokes} examined the different possible terminal sedimentation behaviours allowed for various body symmetries. However, there are bodies which do not attain a steady terminal state.
Recent 
work on sedimentation of a U-shaped disk \cite{miara2024dynamics} shows that achiral bodies can display quasi-periodic dynamics with chiral trajectories, and provides theory for the pitch and roll dynamics. This study also argues that the most complicated motion an arbitrary body can perform is quasi-periodic, and not chaotic. We present a general theory for the sedimentation of any body with two planes of symmetry, henceforth called \textit{di-bilaterals}, and find three kinds of dynamics, the most interesting being quasi-periodic. The body shape allows for apriori classification in sedimentation behaviour. Our choice of dynamical variables brings out a new conserved quantity which restricts the dynamics of di-bilaterals. We explain the quasi-periodicity using Floquet theory. 

Non-spherical particles are often modelled as ellipsoids to account for their orientation degrees of freedom. But ellipsoids sediment with persistent horizontal drift. We show that only a very narrow class of di-bilaterals, which includes ellipsoids, can show persistent drift. Moreover we prove that generic bodies displaying quasi-periodic motion cannot show persistent drift (see Supplemental Material). Our study is a strong indication that di-bilaterals, which can exhibit settling, drifting, helical and quasi-periodic motion, are a far better model of achiral bodies than ellipsoids, capable of explaining a range of sedimentation behaviour. 
\hj{Elastic fibers provide a natural setting for observing di-bilateral sedimentation \cite{PhysRevLett.94.148104}}.

Our interest is in the sedimentation of a single {\it achiral} body in {\it zero background flow}. We opt for di-bilaterals, which support simple translation-rotation coupling (TRC), unlike ellipsoids. We show that all such bodies fall into one of three classes: settlers, drifters and flutterers. Settlers asymptotically align one of their principal axes along gravity and fall vertically, while drifters fall obliquely at constant speed. Flutterers are the most interesting, showing quasi-periodic or periodic, but never chaotic, motion. We show that their motion is completely described by an overlaying of Floquet dynamics in the horizontal, onto periodic dynamics in the vertical direction. 
We design a set of di-bilaterals (see figure 1) and obtain their resistance matrices, which relate the external forces to particle motion, numerically by the boundary integral method.    

In the over-damped limit where we work, gravity is balanced by viscous drag. The Stokesian dynamics is conveniently written in the body-fixed coordinate system of unit vectors $\{\boldsymbol{p}_1, \boldsymbol{p}_2, \boldsymbol{p}_3\}$ (see figure 1), as \cite{happel2012low, thorp2019motion, miara2024dynamics}: 
\begin{equation}
\label{eq:resisMat}
    \begin{pmatrix}
    F_1 \\
    F_2 \\
    F_3 \\
    0 \\
    0 \\
    0 \\
    \end{pmatrix}
    =
    \underbrace{\begin{pmatrix}
    A_1 & 0 & 0 & 0 & 0 & B_{31}\\
    0 & A_2 & 0 & 0 & 0 & 0\\
    0 & 0 & A_3 & B_{13} & 0 & 0\\
    0 & 0 & B_{13} & C_1 & 0 & 0\\
    0 & 0 & 0 & 0 & C_2 & 0\\
    B_{31} & 0 & 0 & 0 & 0 & C_3\\
    \end{pmatrix}}_{\cal R}
    \begin{pmatrix}
    v_1 \\
    v_2 \\
    v_3 \\
    \Omega_1 \\
    \Omega_2 \\
    \Omega_3 \\
    \end{pmatrix},
\end{equation}
where $\boldsymbol{F} = F_1 \boldsymbol{p}_1+ F_2 \boldsymbol{p}_2 + F_3 \boldsymbol{p}_3$ is the external force, and $\boldsymbol{V} = v_1 \boldsymbol{p}_1 + v_2 \boldsymbol{p}_2 + v_3 \boldsymbol{p}_3$ and $\boldsymbol\Omega = \Omega_1 \boldsymbol{p}_1 + \Omega_2 \boldsymbol{p}_2 + \Omega_3 \boldsymbol{p}_3$ the translational and angular velocities respectively. 
Equation \ref{eq:resisMat} has been non-dimensionalized by  characteristic length and time scales $L$ and $\tau=\eta L^2/ (mg_r)$, $\eta$ being the fluid viscosity, and $mg_r$ the body's buoyancy-corrected weight.
The non-dimensional resistance matrix $\cal R$ depends only on the body shape. 
The geometry allows for TRC of a particularly simple form, with two independent entries $B_{13}$ and $B_{31}$. The lab-fixed and body-fixed coordinate axes are related by $\boldsymbol{p}_i = \boldsymbol{R}_{ji}\boldsymbol{e}_j$, $i,j \in \{1,2,3\}$ with $\boldsymbol{e}_1=\boldsymbol{\hat x}, \boldsymbol{e}_2=\boldsymbol{\hat y}$ anti-parallel to gravity, and  $ \boldsymbol{e}_3=\boldsymbol{\hat z}$. So, in the lab frame
\begin{gather}
    \label{eq:Rmat}
    \boldsymbol{R} = 
    \begin{pmatrix}
        p_{1x} && p_{2x} && p_{3x}\\
        p_{1y} && p_{2y} && p_{3y}\\
        p_{1z} && p_{2z} && p_{3z}
    \end{pmatrix}, \quad \boldsymbol{R}^{T}\boldsymbol{R}=\boldsymbol{R}\boldsymbol{R}^T=\mathbb{I}.
\end{gather}
\begin{figure}[h]
\includegraphics[scale=0.23]{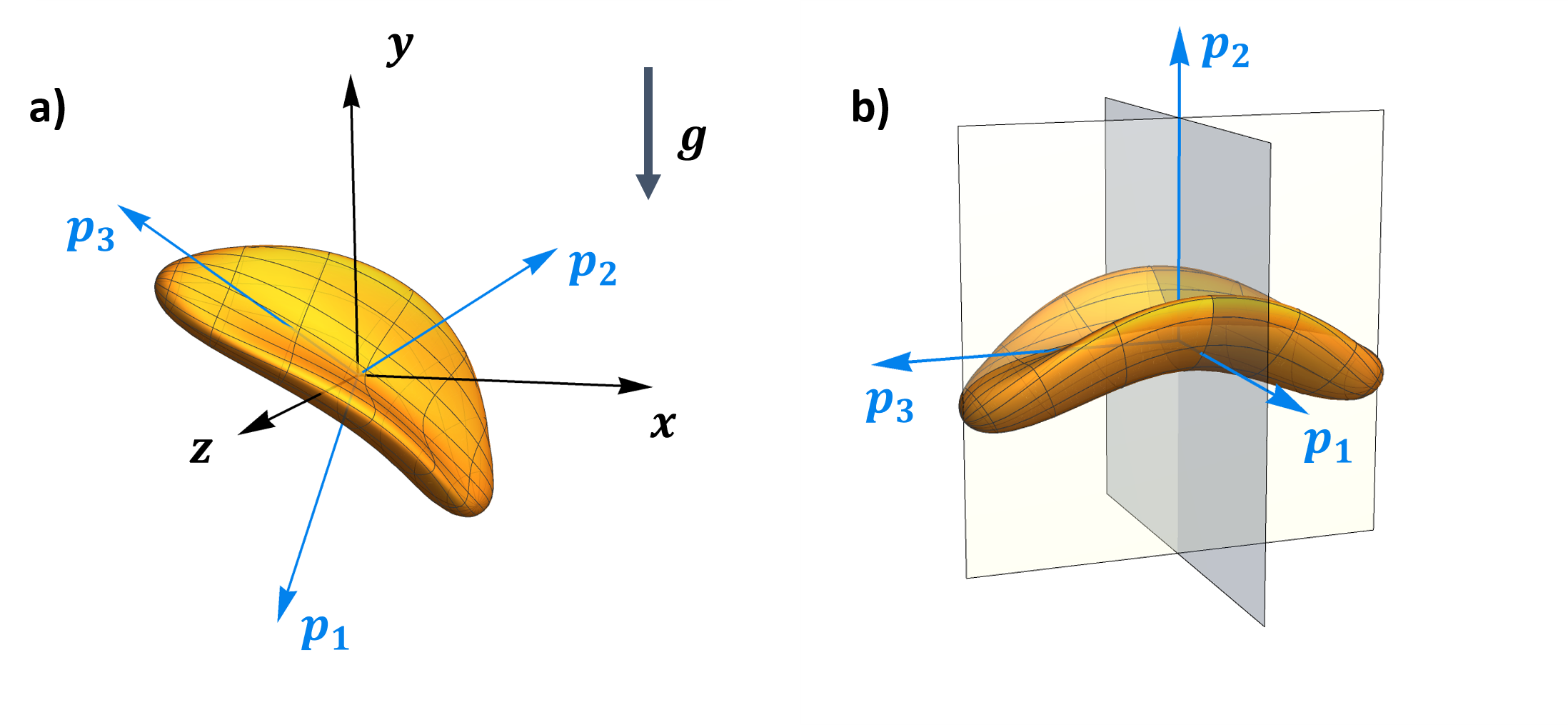}
\caption{\label{fig:bodyOrient} (a) 
A body with two mutually perpendicular planes of symmetry, and the coordinate systems. (b) Showing that $\boldsymbol{p}_1-\boldsymbol{p}_3$ is not a plane of symmetry.}
\end{figure}
We define parameters
\begin{gather}
    \label{eq:alphaParam}
    \alpha_p = \frac{B_{13}}{A_3C_1-B_{13}^2}, \quad
    \alpha_r = \frac{B_{31}}{A_1C_3-B_{31}^2},
\end{gather}
which are functions only of the body shape. 
Rotational invariance in the body frame about the gravity axis provides, for any arbitrary-shaped body, a 
decoupling of the dynamics of the vertical projections of the body's coordinate-axes
from that of their horizontal projections \cite{miara2024dynamics, makino2005sedimentation, moths2013orientational, witten2020review}.
For the vertical projections, using equations \eqref{eq:resisMat}, \eqref{eq:Rmat} and $\boldsymbol{\dot p}_i = \boldsymbol{\Omega}\times \boldsymbol{p}_i,\, i \in \{1,2,3 \}$, we get 
\begin{subequations}
    \label{eq:pyDot}
\begin{gather}
    \dot p_{1y} = \alpha_r p_{1y} p_{2y},\\
    \dot p_{2y} = \alpha_p p_{3y}^2 - \alpha_r p_{1y}^2,\\
    \dot p_{3y} = -\alpha_p p_{3y} p_{2y},
\end{gather}    
\end{subequations}
with the constraint $|\boldsymbol{p}_y|^2  =1$, where $\boldsymbol{p}_y\equiv (p_{1y},p_{2y},p_{3y})$, as required by equation \eqref{eq:Rmat}. Equation \eqref{eq:pyDot} admits a conserved quantity 
\begin{gather}
    \label{eq:Heqn}
    H = |p_{1y}|^{\alpha_p} |p_{3y}|^{\alpha_r}. 
\end{gather}
\hj{The relationship between $H$ and the continuous symmetries of \eqref{eq:pyDot} is discussed in the Supplemental Material \cite{supplemental}.}
Since $p_{1y}=0$ and $p_{3y}=0$ are invariant solutions of \eqref{eq:pyDot}, the signs of $p_{1y}$ and $p_{3y}$ are preserved. We have exploited this to define $H$ to be a positive real number. The conserved quantities $H$ and $|\boldsymbol{p}_y|^2=1$ render the system \eqref{eq:pyDot}, integrable, and restrict the solutions to go towards a fixed point or be periodic. Solutions lie on the intersection of the unit sphere $\mathcal{S}^2$ with surfaces of constant $H$. $H$ is a generalisation to arbitrary di-bilaterals, of a conserved quantity shown by \cite{makino2005sedimentation, witten2020review} for any body whose mobility centre coincides with the centre of mass, which for di-bilaterals is those with $\alpha_p=\alpha_r$.

\begin{figure}[h]
\includegraphics[scale=0.25]{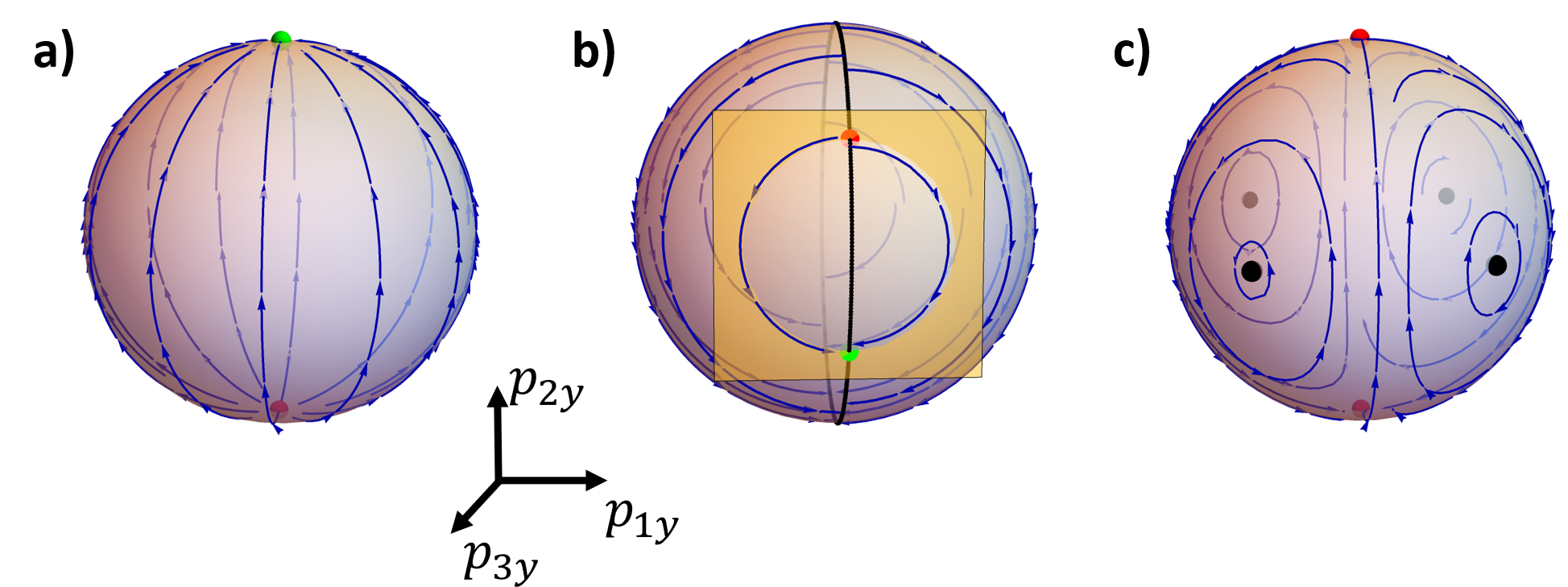}
\caption{\label{fig:phasePlotpY} Phase portraits of the $\boldsymbol{p}_y$ dynamical system. Red, green and black dots denote unstable, stable and centre fixed points, respectively, and blue lines with arrows are sample trajectories. (a) Settlers and (b) drifters ultimately align their $\boldsymbol{p}_2$ axis along and obliquely to gravity respectively. The beige plane in (b) represents a particular $H$. (c) Flutterers rotate forever.}
\end{figure}

Earlier studies \cite{makino2005sedimentation, moths2013orientational,witten2020review} on the possible sedimentation dynamics of bodies of arbitrary shapes focused on the part of the dynamics described by the gravity vector viewed in the body frame. This dynamical quantity is equivalent to our $\boldsymbol{p}_y$. They showed that the $\boldsymbol{p}_y$ dynamics of a generic body can have stable, unstable, saddle and centre fixed points along with limit cycles. Our conserved quantity $H$, removes the possibility of a limit cycle in the $\boldsymbol{p}_y$ dynamics of di-bilaterals, but retains all other possible fixed points, as shown in figure \ref{fig:phasePlotpY}.

Equation \eqref{eq:pyDot} supports the fixed points 
\begin{gather}
    \label{eq:pyFP}
    \boldsymbol{p}_y^* = \{0, \pm 1, 0\}\cup 
    \left\{\pm \sqrt{\frac{\alpha_p}{\alpha_p+\alpha_r}}, 0, \pm \sqrt{\frac{\alpha_r}{\alpha_p+\alpha_r}}\right\},
\end{gather}
and presents three different dynamics for bodies whose $\alpha_p \alpha_r \lesseqgtr 0$, as seen in their phase portraits in figure \ref{fig:phasePlotpY}. 

Di-bilaterals with $\alpha_p \alpha_r<0$ are termed `settlers', whose phase-portrait has one globally attracting (see 
Supplemental Material) stable fixed point and one unstable fixed point. Asymptotically in time, settlers orient their $\boldsymbol{p}_2$ parallel/anti-parallel to gravity and thence fall vertically, without rotating. 
\hj{Similar behaviour occurs in bent achiral fibres \cite{tozzi2011settling} and  non-deformable chains \cite{ekiel2009hydrodynamic}.}
Incidentally, a body lying in this class shows settler-like dynamics at moderate Reynolds number \cite{PhysRevFluids.8.L062301}.

The class we term `drifters' have $\alpha_p \alpha_r = 0$. Their dynamics supports infinitely many fixed points lying along the great circles $\cal A$, of $p_{1y}=0$ if $\alpha_p=0$ or $p_{3y}=0$ if $\alpha_r=0$ (figure \ref{fig:phasePlotpY}(b)). The dynamics occur in the intersection of $\mathcal{S}^2$ with the plane of constant $p_{3y}$ ($p_{1y}$), provided $\alpha_p=0$ ($\alpha_r=0$). This invariant manifold (a circle) intersects $\cal A$, at a stable and an unstable fixed-point. 
Drifters eventually fall with their principal axes inclined at some constant angle to gravity and, like ellipsoids, display persistent horizontal drift and no rotation. 

\begin{figure}[h]
\includegraphics[scale=0.26]{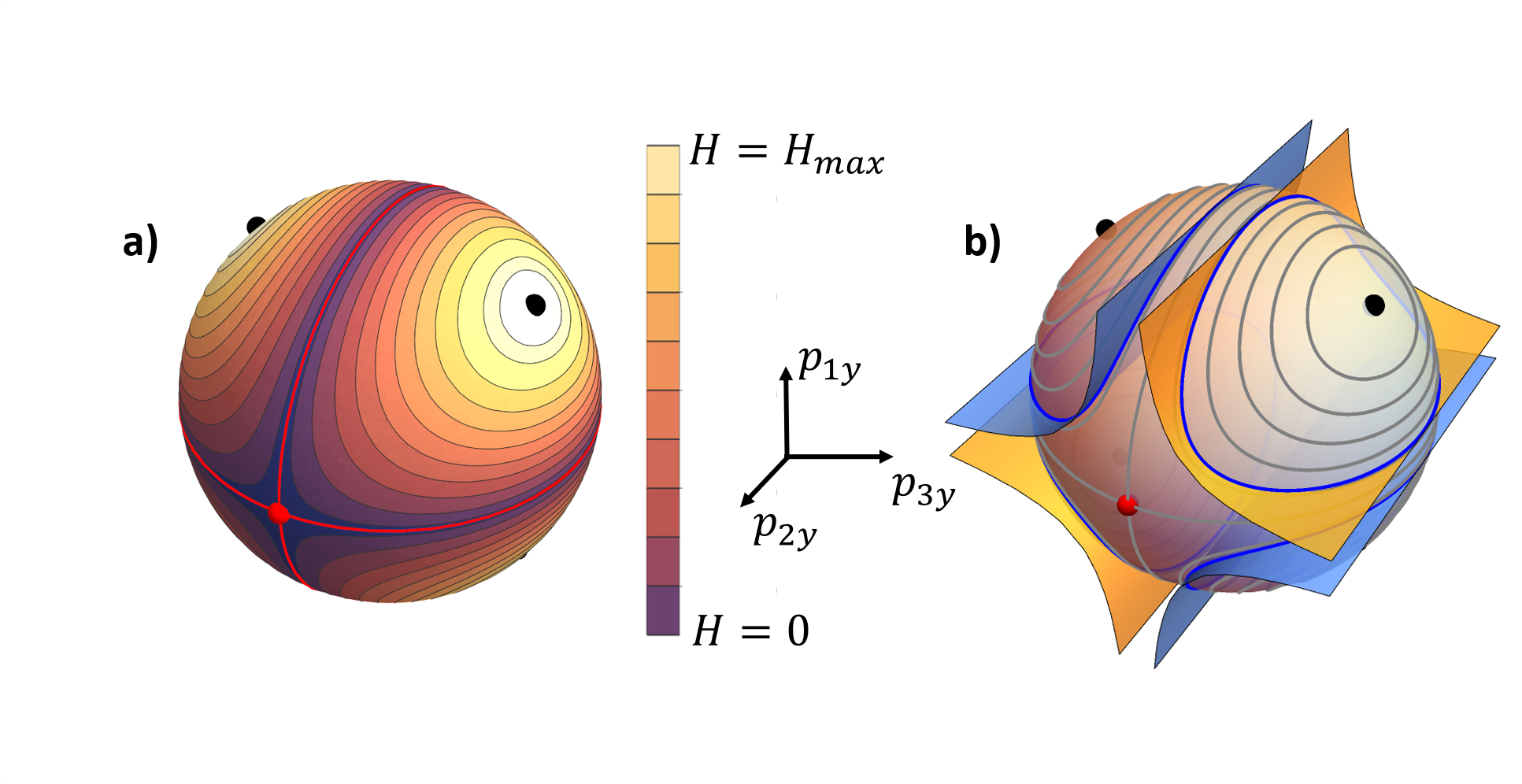}
\caption{\label{fig:Hcurves} (a) Contour plot of the conserved quantity $H$ for flutterers. Closed curves are of constant $H$, with $H_{\text{max}}$ shown by the black dots, and $H=0$ corresponding to the red heteroclinic orbits of the saddle points (red dots). (b) Trajectories on the yellow surfaces with $p_{1y} p_{3y}>0$ have different translational chirality from those with $p_{1y} p_{3y}<0$ on the blue surfaces, as shown in the 
Supplemental Material.}
\end{figure}

\begin{figure}[h]
\includegraphics[scale=0.31]{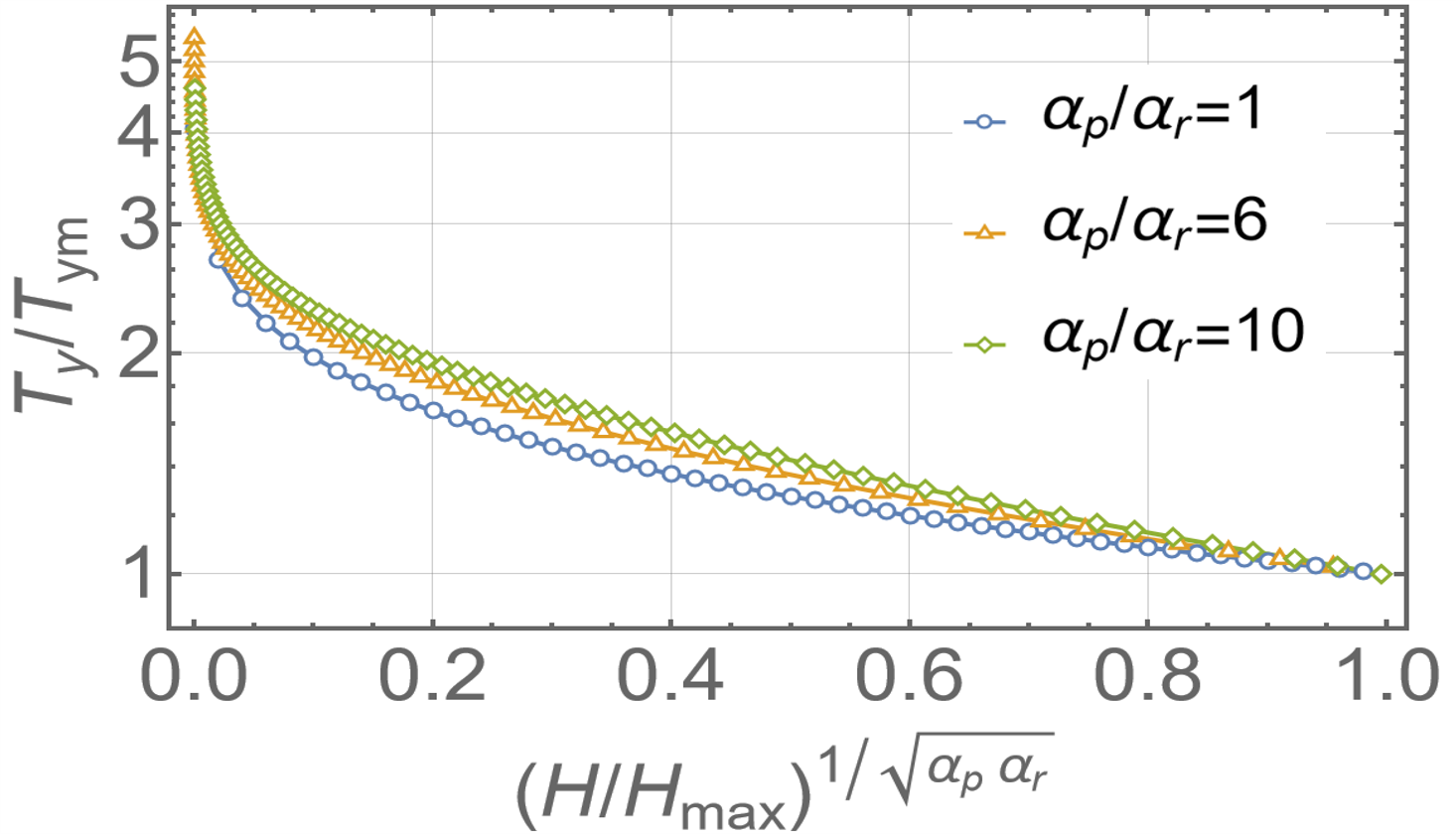}
\caption{\label{fig:TvsH} Time period of the $\boldsymbol{p}_y$ dynamics of flutterers for different $H$. As $H\to 0$, $T_y$ can increase without bound. A given ratio $\alpha_p/\alpha_r$ and its inverse yield the same curve.}
\end{figure}

A range of behaviour, all involving perpetual rotation, is displayed by `flutterers', whose $\alpha_p \alpha_r>0$. Every fixed point here is either a centre or a saddle point, \hj{as for a U-shaped disk \cite{vaquero2024u}}.
The maximum in $H$, 
\begin{gather}
H_{\text{max}} = \dfrac{|\alpha_p|^{\alpha_p/2} |\alpha_r|^{\alpha_r/2}}{|\alpha_p + \alpha_r|^{\alpha_p/2 + \alpha_r/2}},
\end{gather}
corresponds to centre fixed points and the minimum, $H_{\text{min}}=0$, corresponds to stable and unstable manifolds of the saddle points. 
The trajectory for a given initial condition lies on a closed curve representing a constant $H$ surface (figure $\ref{fig:Hcurves}$), 
indicating periodic behaviour of $\boldsymbol{p}_y$. 
A similar periodic behavior in the inclination angles of a U-shaped disk relative to gravity has also been explained in the context of its sedimentation \cite{miara2024dynamics}.
The time period $T_y$ of orbits near the centre fixed points can be obtained upon linearizing equation \eqref{eq:pyDot} about them: $\lim_{H\to H_{\text{max}}}T_y \equiv T_{ym} = 2\pi/\sqrt{2\alpha_p \alpha_r}$. Also $T_y$ everywhere can be obtained numerically, and is shown in figure \ref{fig:TvsH}. Flutterers of different shapes show qualitatively similar trends in time period. Generically, displacing the mobility centre from the centre of mass leads to bifurcation in $\boldsymbol{p}_y$ dynamics \cite{witten2020review, moths2013orientational, GONZALEZ_GRAF_MADDOCKS_2004}. However, such displacement in our flutterers preserves the nature of fixed points.

For the remaining part of the dynamics we write equations for the projection of $\boldsymbol{p}_i$'s along the $x$-axis:
\begin{subequations}    
    \label{eq:pxDot}
\begin{gather}
    \begin{pmatrix}
    \dot p_{1x} \\
    \dot p_{2x} \\
    \dot p_{3x} 
    \end{pmatrix} = 
    \boldsymbol{A}(t)   
    \begin{pmatrix}
     p_{1x} \\
     p_{2x} \\
     p_{3x} 
    \end{pmatrix}, \\
    \boldsymbol{A}(t) \equiv
    \begin{pmatrix} 
     0 && \alpha_r p_{1y}(t) && 0\\
     -\alpha_r p_{1y}(t) && 0 && \alpha_p p_{3y}(t)\\
     0 && -\alpha_p p_{3y}(t) && 0\\
    \end{pmatrix}.
\end{gather}    
\end{subequations}
Note that equation \eqref{eq:Rmat} provides three constraints:
\begin{subequations}
     \label{eq:constraintEqn}    
\begin{gather}
    |\boldsymbol{p}_y(t)|^2=1, \\
    |\boldsymbol{p}_x(t)|^2=1, \\
    \boldsymbol{p}_x(t)\bm{\cdot} \boldsymbol{p}_y(t) = 0,\quad \forall\quad  t.
\end{gather}
\end{subequations}
Equations \eqref{eq:pyDot}, \eqref{eq:pxDot} and \eqref{eq:constraintEqn} fully describe the rotational dynamics of 
di-bilaterals. Equation \eqref{eq:pxDot} shows that $\boldsymbol{p}_x$ is driven by $\boldsymbol{p}_y$. Since the latter is time-periodic for flutterers, $\boldsymbol{p}_x(t)$ can be solved using Floquet theory \cite{chicone2006ordinary}. 
We define a $T_y$ period Poincare map of the $\boldsymbol{p}_x$ dynamical system as:
\begin{subequations}
\label{eq:pMapDef}
\begin{gather}
    \mathcal{P}_{t_0}^{t_0+T_y} = \boldsymbol{\Phi}_{t_0}^{t_0+T_y}, \quad \\
    \boldsymbol{\dot \Phi}_{t_0}^{t} = \boldsymbol{A}(t)\boldsymbol{\Phi}_{t_0}^{t};\quad \boldsymbol{\Phi}_{t_0}^{t_0} = \mathbb{I},
\end{gather}    
\end{subequations}
where $\boldsymbol{\Phi}_{t_0}^t$ is the solution operator of  equation \eqref{eq:pxDot}, i.e., $\boldsymbol{p}_x(t) = \boldsymbol{\Phi}_{t_0}^t\boldsymbol{p}_x(t_0)$. Since $\boldsymbol{A}^T(t) = -\boldsymbol{A}(t)$, equation \eqref{eq:pMapDef} ensures that $\boldsymbol{\Phi}_{t_0}^t \in SO(3)$. Therefore, both the solution operator and the Poincare map correspond to some 3D rotations. Consequently, the eigenvalues of the Poincare map are given by $1$ and $e^{\pm i \mu T_y}$, where $ \mu \in \mathbb R$ is the Floquet exponent. 
The constraints of \eqref{eq:constraintEqn} restrict $\boldsymbol{p}_x(t)$ to lie normal to $\boldsymbol{p}_y(t)$. Since $\boldsymbol{p}_y(t_0+T_y) = \boldsymbol{p}_y(t_0)$, the image set of the Poincare map $\{\mathcal{P}_{t_0}^{t_0+nT_y}\boldsymbol{p}_x(t_0)\}_{n\in \mathbb N}$ is a subset of the great circle lying normal to $\boldsymbol{p}_y(t_0)$. Thus, the Poincare map $\mathcal{P}_{t_0}^{t_0+T_y}$ acting on $\boldsymbol{p}_x$ corresponds to the rotation of $\boldsymbol{p}_x$ by an angle $\theta\equiv \mu T_y$ about the axis $\boldsymbol{p}_y(t_0)$.
The Supplemental Material shows that $\theta$ depends only on $H$ and not on $t_0$.
By Floquet theory, any solution of equation \eqref{eq:pxDot} can be decomposed as 
\begin{gather}
    \label{eq:pxDecomp}
    \boldsymbol{p}_x(t) = e^{i\mu t}\boldsymbol{b}(t),
\end{gather}
where $\boldsymbol{b}(t) = \boldsymbol{b}(t+T_y)$ is some $T_y$-periodic function. Thus, if the driving frequency $\omega_y \equiv 2\pi/T_y$ is incommensurate with the response frequency $\mu$, i.e., if $\theta/2\pi = \mu/\omega_y$ is an irrational number, flutterers' motion is quasi-periodic, whereas if it is rational we have periodic dynamics. In the quasi-periodic case the image set of the Poincare map fills up the entire great circle.

\begin{figure}[h]
\includegraphics[scale=0.26]{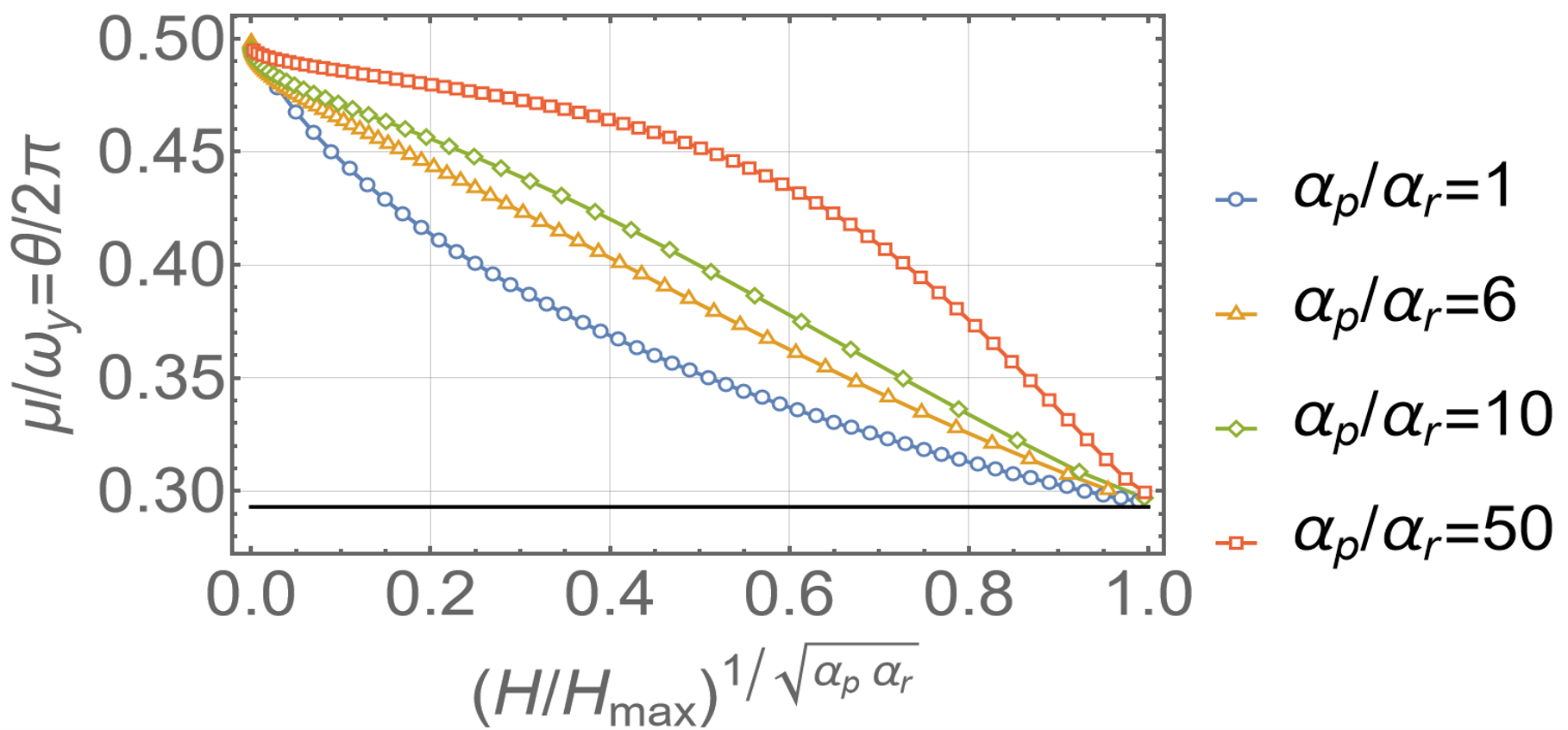}
\caption{\label{fig:freqRatio} The $\boldsymbol{p}_x$ dynamics of flutterers involves two frequencies, $\mu$ and $\omega_y$. Their ratio, shown here, is the rotation angle $\theta/2\pi$ of the Poincare map. The black horizontal line is the lower limit $1-1/\sqrt{2}$ of the ratio.}
\end{figure}

\begin{figure}[h]
\includegraphics[scale=0.31]{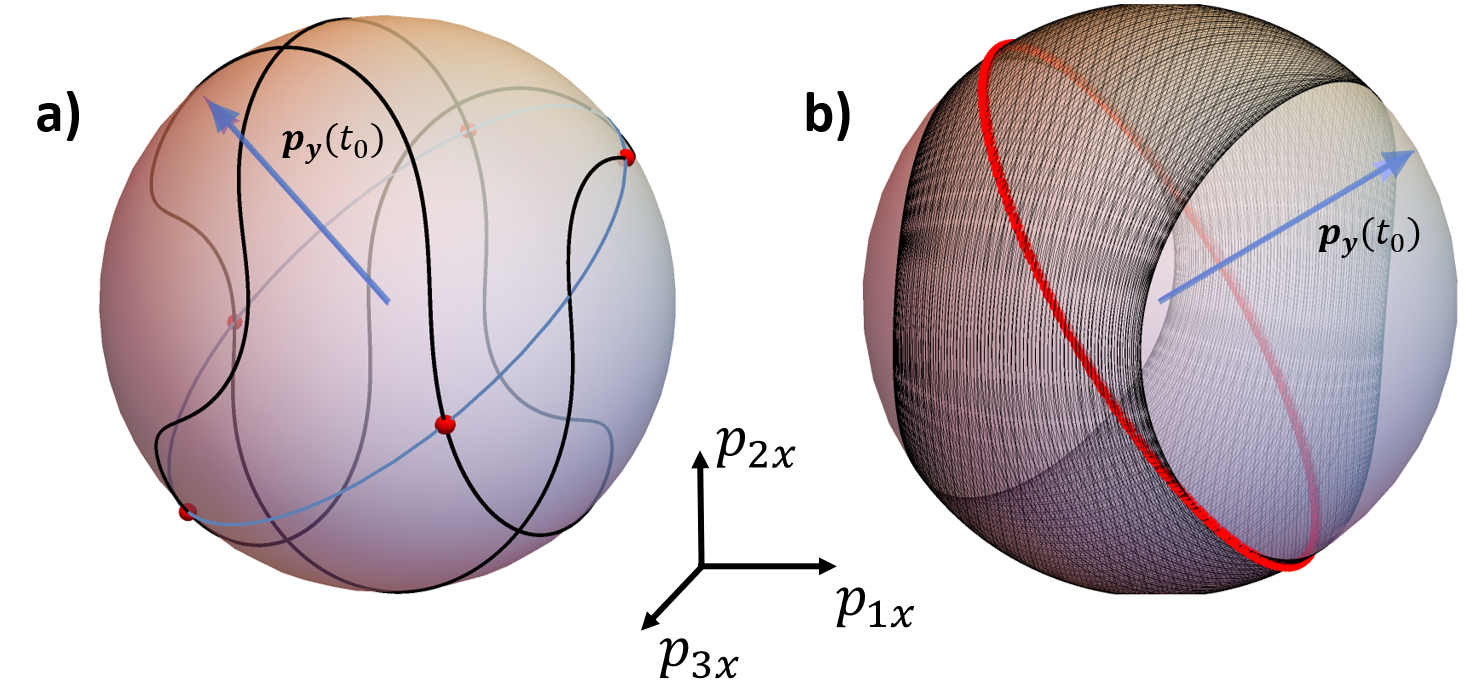}
\caption{\label{fig:pxTraj} Sample $\boldsymbol{p}_x$ trajectories shown as black curves on $\mathcal{S}^2$. (a) Period-5 trajectory. The red dots show the Poincare map $\mathcal{P}_{t_0}^{t_0+T_y}$ which lie on the great circle shown by the blue curve. This great circle lies normal to $\boldsymbol{p}_y(t_0)$. (b) Quasi-periodic trajectory. The Poincare map covers the entire great circle lying normal to $\boldsymbol{p}_y(t_0)$. The filled black region has the inversion symmetry $\boldsymbol{p}_x \to -\boldsymbol{p}_x$. }
\end{figure}

Figure \ref{fig:freqRatio} shows the dependence of the frequency ratio on $H$. 
There are differences in detail for different flutterer shapes, but in all cases, the frequency ratio varies continuously on $[1-1/\sqrt{2},\, 0.5]$, showing dense existence of periodic and quasi-periodic orbits. Figure \ref{fig:pxTraj} shows phase trajectories and image sets of Poincare maps for a periodic and quasi-periodic case.

Another class of periodic orbits are obtained on the centre fixed points of the $\boldsymbol{p}_y$ dynamics, characterized by $H=H_{\text{max}}$. In this case $\boldsymbol{A}(t) = \boldsymbol{A}^*$ is a constant matrix with the eigenvalues $0, \pm i \sqrt{\alpha_p \alpha_r}$. The eigenvector corresponding to the zero eigenvalue is not orthogonal to $\boldsymbol{p}_y$, and is dropped from the discussion since it violates the constraints. In this case, the body simply rotates about gravity and falls in a helical trajectory \cite{witten2020review}.

The translational velocity of di-bilaterals in the lab frame depends directly on $\boldsymbol{p}_x$ and $\boldsymbol{p}_y$ (see Supplemental Material). Thus, the translational trajectories inherit the periodicity or quasi-periodicity from the angular dynamics.
A flutterer does not display persistent horizontal drift and maintains the same sign of chirality (see figure \ref{fig:xzTraj}) throughout (more details in the Supplemental Material).

\begin{figure}[h]
\includegraphics[scale=0.29]{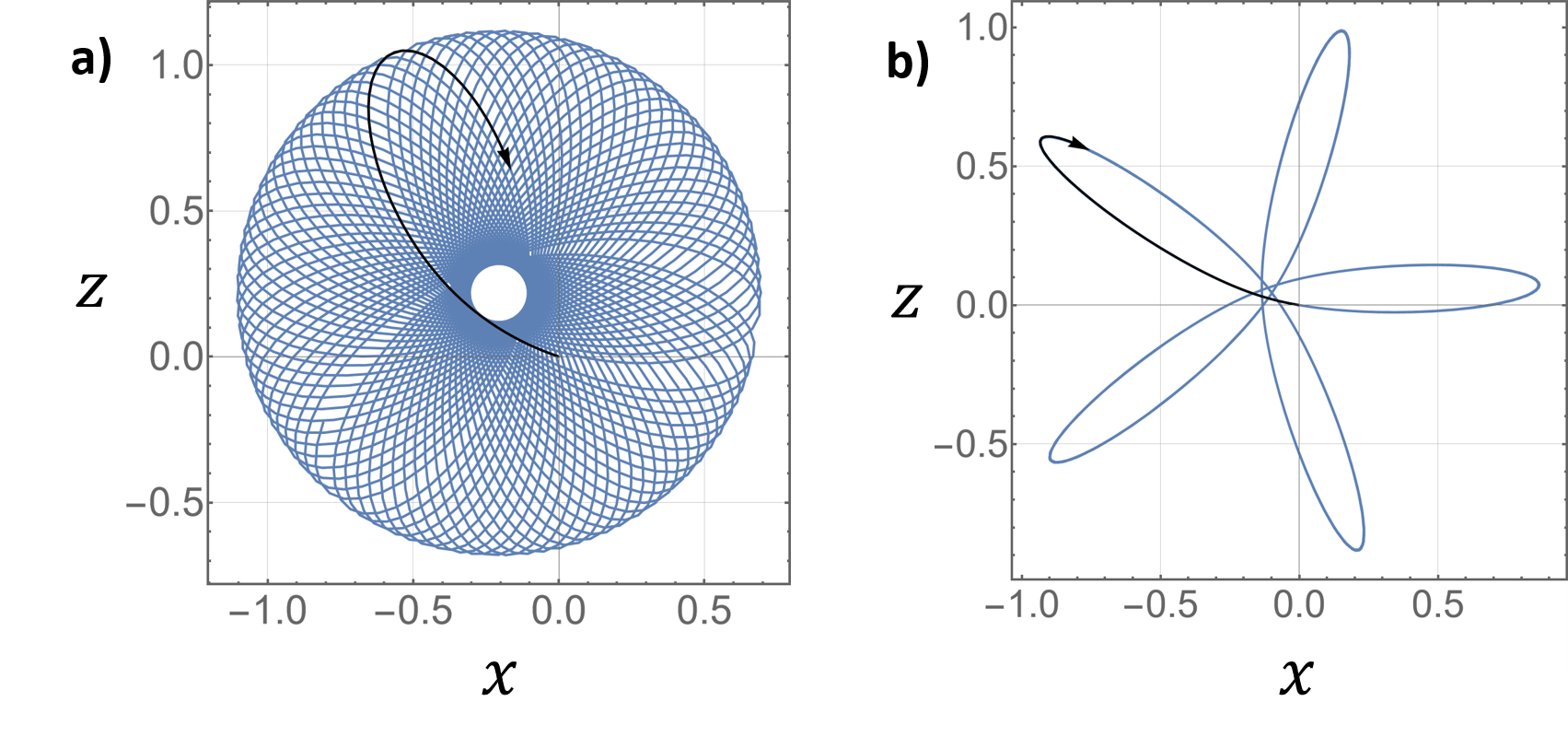}
\caption{\label{fig:xzTraj} Sample trajectories 
in the horizontal plane for (a) quasi-periodic and (b) periodic orbits. The black arrows indicate the chirality of the settling trajectories.
} 
\end{figure}


It remains to obtain the resistance matrix $\cal R$, and thence $\alpha_p$ and $\alpha_r$, for a given 
di-bilateral. We design a set of bodies with two parameters: the bending parameter $a$, and the concavity parameter $b$, as shown in figure \ref{fig:bodyDescptn} (details in section 5 of Supplemental Material).
\begin{figure}[h]
\includegraphics[scale=0.22]{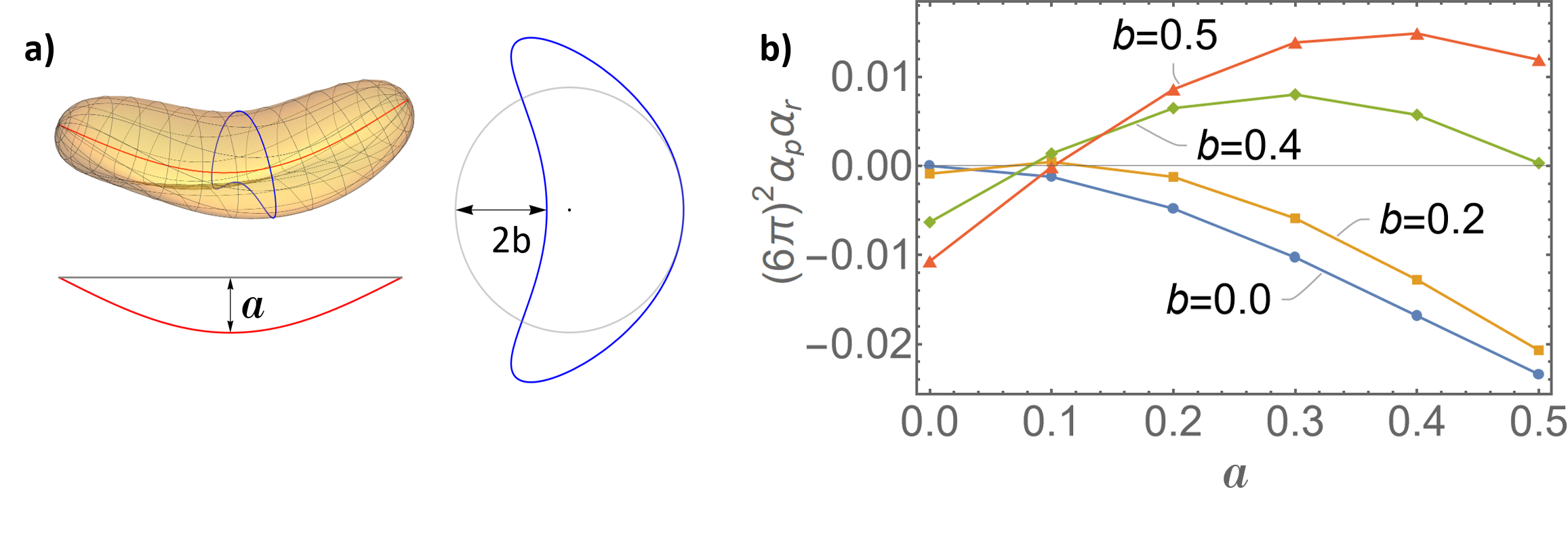}
\caption{\label{fig:bodyDescptn} a) A set of bodies is defined by the parameters $a$ and $b$. b) As $a$ and $b$ defined are varied, we obtain settlers ($\alpha_p \alpha_r<0$), drifters ($\alpha_p \alpha_r=0$), and flutterers ($\alpha_p \alpha_r>0$). Increasing $b$ turns settlers into flutterers, but increasing $a$ at fixed $b$ eventually turns flutterers back into settlers.}
\end{figure}
We use the Boundary Integral Method \cite{kim2013microhydrodynamics,bagge2021highly,pozrikidis1992boundary,pozrikidis2002practical, prosperetti2009computational} to obtain $\cal R$ numerically. The parameters $\alpha_p,\, \alpha_r,\, G_1,\, G_2,\, G_3$ completely describe the sedimentation problem in a quiescent fluid, with figure \ref{fig:bodyDescptn} showing their dependence on shape.
\hj{For a given $b$, an optimal  $a$ maximizes $\alpha_p\alpha_r$, resulting in the highest fluttering frequency.}

To summarize, we have shown that 
di-bilaterals show three kinds of Stokesian sedimentation, depending only on their shape, appearing through the parameter $\alpha_p \alpha_r$. Settlers and drifters asymptotically fall vertically and obliquely respectively, without rotating. In flutterers, periodic dynamics in $\boldsymbol{p}_y$ drives a periodic or quasi-periodic response of $\boldsymbol{p}_x$, based on the ratio $\mu/\omega_y$. Both time-scales depend only on the conserved quantity $H$.
All three classes of bodies can be obtained by bending a capsule about its central axis and making its cross-section concavo-convex.
Previous studies on asymmetric bodies have well-understood $\boldsymbol{p}_y$ dynamics, predicting fixed points of differing character in  \cite{GONZALEZ_GRAF_MADDOCKS_2004, makino2005sedimentation, moths2013orientational, witten2020review}. Among these, we rule out limit cycles for di-bilaterals by additional conserved quantity, $H$. Except for a qualitative argument which argues for of non-chaotic dynamics \cite{miara2024dynamics}, the  $\boldsymbol{p}_x$ dynamics has not been studied before to our knowledge.
The $\boldsymbol{p}_x$ dynamics enables an understanding and quantitative description of quasi-periodicity. We also extend the proof  \cite{makino2005sedimentation} of no persistent drift of torque-free skewed bodies to any arbitrary body which performs periodic/quasi-periodic motion.

\begin{acknowledgments}
We thank Saumav Kapoor for help in designing flutterers. We acknowledge support of the Department of Atomic Energy, Government of India, under project no. RTI4001. We thank the referees for their suggestions, which improved the paper.
\end{acknowledgments}

\appendix

\section{Lyapunov function for global stability}
\label{app:fLP}

For settlers, which have $\alpha_r \alpha_p<0$, a Lyapunov function exists which shows that the stable fixed point is globally stable. 

Consider first the case where $\alpha_p>0$ and $\alpha_r<0$: here $\boldsymbol{p}_y^{*1} = (0, 1, 0)$ is a non-hyperbolic fixed point corresponding to the eigenvalues $(0, -\alpha_p, -|\alpha_r|)$. The Lyapunov function is
\begin{gather}
    \label{eq:LPfn1}
    f_1(\boldsymbol{p}_y) = p_{1y}^2+p_{3y}^2 + (p_{2y}-1)^2.
\end{gather}
Clearly, $f_1(\boldsymbol{p}_y^{*1}) = 0$ and $f_1(\boldsymbol{p}_y)>0$ for $\boldsymbol{p}_y\neq \boldsymbol{p}_y^{*1}$. Now, $\dot f_1 = -2(\alpha_p p_{3y}^2 + |\alpha_r| p_{1y}^2) < 0$ for all $\boldsymbol{p}_y\neq \boldsymbol{p}_y^{*1}$, showing  $\boldsymbol{p}_y^{*1}$ to be a  globally attracting fixed point.

Similar arguments may be made for the other case, where $\alpha_p<0$ and $\alpha_r>0$, using the Lyapunov function
\begin{gather}
    \label{eq:LPfn2}
    f_2(\boldsymbol{p}_y) = p_{1y}^2+p_{3y}^2 + (p_{2y}+1)^2,
\end{gather}
with $\boldsymbol{p}_y^{*1} = (0, -1, 0)$ being the stable fixed point.

\section{Discrete and continuous symmetries of the rotational dynamics of flutterers}

The discrete symmetry group of the rotational dynamical equations (4) and (8) in the main paper is $\mathcal{G}_R=\mathbb{Z}_2 \times \mathbb{Z}_2 \times \mathbb{Z}_2$. The group $\mathcal{G}_R$ is an abelian group containing 8 elements which corresponds to the following eight transformations of dynamical variables:
\begin{table}[h]
    \centering
    \begin{tabular}{|c|c|}
    \hline
    Element & Action on ($p_{1y},\, p_{2y},\, p_{3y},\, p_{1x},\, p_{2x},\, p_{3x}$)  \\
    \hline
    $e$ & ($p_{1y},\, p_{2y},\, p_{3y},\, p_{1x},\, p_{2x},\, p_{3x}$) \\
    $a$ = $g \ast h$ & ($-p_{1y},\, p_{2y},\, -p_{3y},\, p_{1x},\, -p_{2x},\, p_{3x}$) \\
    $b$ = $a\ast c$ & ($-p_{1y},\, p_{2y},\, p_{3y},\, p_{1x},\, -p_{2x},\, -p_{3x}$) \\
    $c$ = $a\ast b$ & ($p_{1y},\, p_{2y},\, -p_{3y},\, p_{1x},\, p_{2x},\, -p_{3x}$) \\
    $d$ = $a\ast f$ & ($p_{1y},\, p_{2y},\, -p_{3y},\, p_{1x},\, -p_{2x},\, -p_{3x}$) \\
    $f$ = $b\ast h$ & ($-p_{1y},\, p_{2y},\, p_{3y},\, -p_{1x},\, p_{2x},\, p_{3x}$) \\
    $g$ = $a\ast h$ & ($-p_{1y},\, p_{2y},\, -p_{3y},\, -p_{1x},\, p_{2x},\, -p_{3x}$) \\
    $h$ = $c\ast d$ & ($p_{1y},\, p_{2y},\, p_{3y},\, -p_{1x},\, -p_{2x},\, -p_{3x}$) \\
    \hline
    \end{tabular}
    \caption{Discrete symmetries of rotational dynamics}
    \label{tab:discreteSym}
\end{table}

The binary operations between the elements is shown in the first column of table \ref{tab:discreteSym}. The corresponding multiplication table of this group is isomorphic to the group $\mathbb{Z}_2 \times \mathbb{Z}_2 \times \mathbb{Z}_2$. The group element $h$ is used to establish that there is no persistent drift for quasi-periodic orbits, while the element $g$ is used to show the dependence of the chirality on the sign of $p_{1y}p_{3y}$. Here we have restricted ourselves to transformations that only act on the dynamical variables and not on time $t$.

The continuous Lie symmetries of the rotational dynamics are simply scaling symmetries and time translational symmetries. The linearly independent generators of these continuous symmetries are given by
\begin{subequations}
    \label{eq:generators}
\begin{gather}
    X_1 = \boldsymbol{p}_y\cdot\boldsymbol{\nabla}_{\boldsymbol{p}_y} - t\partial_t, \\
    X_2 = \boldsymbol{p}_x\cdot\boldsymbol{\nabla}_{\boldsymbol{p}_x}, \\
    X_3 = \boldsymbol{\dot p}_y \cdot\boldsymbol{\nabla}_{\boldsymbol{p}_y} + \boldsymbol{\dot p}_x \cdot\boldsymbol{\nabla}_{\boldsymbol{p}_x}, \\
    X_4 = \partial_t.
\end{gather}    
\end{subequations}
The corresponding Lie algebra of these generators is:
\begin{subequations}
    \label{eq:LieAlgbra}
    \begin{gather}
        [X_1,\, X_2] = 0, \\
        [X_1,\, X_3] = X_3,\quad [X_1,\, X_4] = X_4, \\
        [X_2,\, X_3] = [X_2,\, X_4] = [X_3,\, X_4] = 0. 
        \end{gather}
\end{subequations}
The generators $X_1$ and $X_2$ correspond to scaling symmetries: $e^{\lambda X_1}(\boldsymbol{p}_y, \boldsymbol{p}_x, t) = (e^{\lambda}\boldsymbol{p}_y, \boldsymbol{p}_x, e^{-\lambda}t)$ and $e^{\lambda X_2}(\boldsymbol{p}_y, \boldsymbol{p}_x, t) = (\boldsymbol{p}_y, e^{\lambda}\boldsymbol{p}_x, t)$ while the other two generators correspond to time translation. These Lie symmetries can be used to construct the conserved quantities or the first integrals of the rotational dynamics, as described in \cite{hydon2000symmetry}. 
The four generators with three unique transformations give three unique characterstics as:
\begin{subequations}
    \label{eq:Qeqn}
    \begin{gather}
    \boldsymbol{Q}_1 = (\boldsymbol{p}_y + t \boldsymbol{\dot p}_y,\, t \boldsymbol{\dot p}_x),\\
    \boldsymbol{Q}_2 = (0,\, 0,\, 0,\, \boldsymbol{p}_x),\\
    \boldsymbol{Q}_3 = (\boldsymbol{\dot p}_y,\, \boldsymbol{p}_x).
    \end{gather}
\end{subequations}
The three co-characterstics are given by: 
\begin{subequations}
    \label{eq:Lmdaeqn}
    \begin{gather}
    \boldsymbol{\Lambda}_1 = (\boldsymbol{p}_y,\, 0,\, 0,\, 0),\\
    \boldsymbol{\Lambda}_2 = (\alpha_p/p_{1y},\, 0,\, \alpha_r/p_{3y},\, 0,\, 0,\, 0),\\
    \boldsymbol{\Lambda}_3 = (0,\, 0,\, 0,\, \boldsymbol{p}_x).
    \end{gather}
\end{subequations}
The conserved quantity $H$ is obtained as a first integral using: $\boldsymbol{Q}_1 \cdot \boldsymbol{\Lambda}_2 = \alpha_p + \alpha_r + t \dfrac{d}{dt}\log( p_{1y}^{\alpha_p} p_{3y}^{\alpha_r} )$. The other scalar products of $\boldsymbol{Q}$'s and $\boldsymbol{\Lambda}$'s yield the constraints (9) in the main paper as the conserved quantities.

\section{Details of dynamics of flutterers}

\subsection{Persistent horizontal drift}
\label{sec:PDFlutterer}
In this section we ask if a di-bilateral, which is a flutterer, can ever show persistent drift in the horizontal plane and answer the question separately for the quasiperiodic and the periodic cases. For the former, following Thorp \& Lister \cite{thorp2019motion} we exploit the discrete symmetries of the rotational dynamical equations to rule out persistent drift. \\
\label{app:NoDriftProof}
\textit{Claim:} \textit{Flutterers can have persistent drift in the plane perpendicular to  gravity only if the driving frequency $\omega_y$ of the $\boldsymbol{p}_y$ dynamics is exactly equal to the response frequency $\mu$ of the $\boldsymbol{p}_x$ dynamics, i.e., $\mu=\omega_y$. In this case, the Poincare map $\mathcal{P}_{t_0}^{t_0+T_y}$ is just the identity map.}
\newline
\textit{Proof:}
The rotational dynamical equations (4) and (8) in the main paper have a discrete symmetry group $\mathcal{G}_R=\mathbb{Z}_2\times \mathbb{Z}_2\times \mathbb{Z}_2$. If a given rotational trajectory is invariant under $g\in \mathcal{G}_R$ and $g(v_i)=-v_i$ then there can be no persistent drift along the $x_i$ direction for that trajectory, for the drift at one point is exactly cancelled by the other mapped point. 
Note that $g:(p_{1y}, p_{2y}, p_{3y}, p_{1x}, p_{2x}, p_{3x}) \to (p_{1y}, p_{2y}, p_{3y}, -p_{1x}, -p_{2x}, -p_{3x})$ is a symmetry of equations (4) and (8) in the main paper with $g(v_x) = -v_x$ and $g(v_z) = -v_z$. Thus the invariance of trajectories of the rotational dynamical system under $g$ is sufficient to rule out any persistent horizontal drift. 

For quasi-periodic cases, any point $\boldsymbol{p}_x(t)$ in a given trajectory will be mapped to the point $-\boldsymbol{p}_x$ at time $t+kT_y$, for some $k \in \mathbb N$. This is due to the fact that the Poincare map $\mathcal{P}_{t}^{t+T_y}$ fills up the entire great circle with normal vector $\boldsymbol{p}_y(t)$. Since this great circle has an inversion symmetry, there exist a $k\in \mathbb N$ such that $\boldsymbol{p}_x(t+kT_y) = -\boldsymbol{p}_x$. Therefore, the region occupied by quasi-periodic orbits on $S^2$ has inversion symmetry $\boldsymbol{p}_x \to -\boldsymbol{p}_x$. This rules out any persistent drift in the horizontal plane owing to the discrete symmetry element $g\in \mathcal{G}_R$. 

For the periodic case we have closed trajectories subject to constraints (9) in the main paper which may not have the inversion symmetry $\boldsymbol{p}_x\to -\boldsymbol{p}_x$ (see figure (6) in the main paper). The discrete symmetry argument to rule out persistent drift is inconclusive in this case. A necessary condition for persistent drift can be obtained as follows: owing to equation (11) in the main paper, each term on the right hand side of $v_x$ and $v_z$ in equation (12) in the main paper is of the form $e^{i\mu t} f(t)$, where $f(t) = f(t+T_y)$ is some $T_y$ periodic function. The contribution of such terms to the displacement over one time period $T$ of a periodic trajectory is
\begin{gather}
    \label{eq:xTval}
    \Delta(T) \equiv \frac{1}{T}\int_0^T e^{i\mu t} f(t)\, dt 
    = \frac{1}{T}\sum_{k \in \mathbb{Z}}\hat f_k\int_0^T e^{i(\mu+\omega_y k) t} \, dt,
\end{gather}
where we used the Fourier expansion: $f(t) = \sum_{k\in\mathbb{Z}} \hat{f}_k e^{i\omega_y t}$ with $\omega_y T_y = 2\pi$.
Now, for the periodic case, $\mu/\omega_y = n_1/n_2 \in \mathbb{Q}$, where $n_1, n_2 \in \mathbb{N}$ and $T=2\pi n_2/\omega_y = 2\pi n_1/\mu$. Using these in equation \eqref{eq:xTval}, we have
\begin{gather}
    \label{eq:xTval2}
    \Delta(T) =  \sum_{k \in \mathbb{Z}}\hat f_k\int_0^1 e^{2\pi i (n_1 + k n_2 )q} \, dq =  \sum_{k \in \mathbb{Z}}\hat f_k \delta_{n_1+kn_2, 0},
\end{gather}
where $q = t/T$. Since $k\in \mathbb{Z}$, the contribution can be non-zero if $n_2=1$ which implies $\mu/\omega_y \in \mathbb{N}$. Now, $\mu = k \omega_y$ for $k\in\mathbb N$ implies $\boldsymbol{p}_x(t+T_x) = \boldsymbol{p}_x(t)$ with $T_x = T_y/k$. But $\boldsymbol{\dot p_x}(t+T_x) = \boldsymbol{\dot p_x}(\boldsymbol p_y(t+T_x), \boldsymbol{p}_x(t)) \neq \boldsymbol{\dot p}_x(\boldsymbol p_y(t), \boldsymbol p_x(t))$, unless $T_y=T_x$. Thus, the $\boldsymbol{p}_x$ trajectory cannot repeat itself after $T_x$ since the tangent vector $\boldsymbol{\dot p}_x$ at time $t+T_x$ is different from what it was at time $t$, unless $k=1$. Therefore, the only periodic trajectory for which $\Delta(T)$ can be non-zero is the case $\omega_y=\mu$.

In section \ref{sec:PD} we show how persistent drift can be ruled out for arbitrary bodies which performs periodic/quasi-periodic motion, except for the special case where the time period of $\boldsymbol{p}_y$ dynamics exactly matches that of the $\boldsymbol{p}_x$ dynamics.

For flutterers, numerics suggests that the ratio $\mu/\omega_y$ is bounded between the interval $[1-1/\sqrt{2},\, 0.5]$ in which case there is no persistent drift. The lower limit of the ratio can be obtained by linearizing $\boldsymbol{p}_y$ dynamics near $H=H_{\text{max}}$, in which case the time period of $\boldsymbol{p}_y$ dynamics is periodic with time period $T_{ym} = 2\pi/\sqrt{2\alpha_p \alpha_r}$. The $\boldsymbol{p}_x$ dynamics upto the zeroth order in $H-H_{\text{max}}$ is given by $\boldsymbol{\dot p}_x = \boldsymbol{A}^*\boldsymbol{p}_x$, where $\boldsymbol{A}^*$ is the matrix in equation (8) of the main paper, evaluated at one of the centre fixed points of the $\boldsymbol{p}_y$ dynamics. Therefore, upto the zeroth order the $\boldsymbol{p}_x$ dynamics is also periodic with the time period $T_{xm} = 2\pi/\sqrt{\alpha_p \alpha_r}$. Now, using equation (11) of the main paper, $\boldsymbol{p}_x(t+T_{ym}) = \boldsymbol{p}_x(t)e^{i \mu T_{ym}}$; and since $\boldsymbol{p}_x(t) \sim e^{\pm 2\pi i t/T_{xm}}$, we have $e^{\pm 2\pi i T_{ym}/T_{xm}} = e^{i\mu T_{ym}}$, which gives $\mu T_{ym}/2\pi = 1 - 1/\sqrt{2}$. 

When $H=H_{\text{max}}$, we still have no persistent drift, since the valid solutions of $\boldsymbol{p}_x: \boldsymbol{p}_x(t) = e^{\boldsymbol{A}^* t}\boldsymbol{p}_x(0)$ are periodic with zero mean. This is because the eigenvector of $\boldsymbol{A}^*$ with zero eigenvalue does not satisfy the constraint $\boldsymbol{p}_x\cdot\boldsymbol{p}_y =0$ and thus cannot be part of the solution. 

Finally, for $H=0$, the body exponentially reaches one of the saddle points $\boldsymbol{p}_y^* = (0,\pm 1, 0)$ along its stable manifold as $t\to \infty$ and consequently the horizontal drift goes to zero exponentially.

\subsection{Chirality of translational trajectories}

We discuss the chirality of the translational trajectories of flutterers. As the body settles, its motion in the horizontal $x-z$ plane is either clockwise or anti-clockwise when viewed from below, corresponding to negative or positive chirality respectively. In figure 3 of the main paper, regions on $\mathcal{S}^2$ with $p_{1y}p_{3y}>0$ correspond to different chirality in translation motion from regions with $p_{1y}p_{3y}<0$. This can be seen as follows: for any given angular trajectory ($\boldsymbol{p}_y, \boldsymbol{p}_x$) with $p_{1y}p_{3y}>0$, the transformation $g:(p_{1y},\, p_{1x}) \to (-p_{1y},\, -p_{1x})$ maps the original trajectory to the trajectory in the region $p_{1y}p_{3y}<0$ with the same $H$. Thus the mapped trajectory has same time period $T_y$ in $\boldsymbol{p}_y$ dynamics and same frequency ratio $\mu/\omega_y$ in $\boldsymbol{p}_x$ dynamics. However, $g:v_z \to -v_z$, while the other velocity components remain invariant. So the trajectory of the translational dynamics ($x(t), y(t), z(t)$) is mapped to the trajectory $(x(t), y(t), -z(t))$, i.e., the original and mapped trajectories are of opposite chirality.



\section{Similarity between Poincare maps with different reference times}
Consider two Poincare maps $\mathcal{P}_{t_1}^{t_1+T_y}$ and $\mathcal{P}_{t_2}^{t_2+T_y}$ with $t_2>t_1$ and $T_y$ the same in both the maps. These map satisfies the following equation:
\begin{subequations}
    \label{eq:Poincare2maps}
    \begin{gather}
        \mathcal{P}_{t_1}^{t_1+T_y} = \boldsymbol{\Phi}_{t_1}^{t_1+T_y};\, \boldsymbol{\dot\Phi}_{t_1}^{t_1+t} = \boldsymbol{A}(t)\boldsymbol{\Phi}_{t_1}^{t_1+t},\, \boldsymbol{\Phi}_{t_1}^{t_1} = \mathbb{I}, \\
        \mathcal{P}_{t_2}^{t_2+T_y} = \boldsymbol{\Phi}_{t_2}^{t_1+T_y};\, \boldsymbol{\dot\Phi}_{t_2}^{t_2+t} = \boldsymbol{A}(t)\boldsymbol{\Phi}_{t_2}^{t_2+t},\, \boldsymbol{\Phi}_{t_2}^{t_2} = \mathbb{I}.
    \end{gather}
\end{subequations}
\textit{Claim:} The two Poincare maps $\mathcal{P}_{t_1}^{t_1+T_y}$ and $\mathcal{P}_{t_2}^{t_2+T_y}$ are related by a similarity transformation
\newline
\textit{Proof:} 
Since $\boldsymbol{\Phi}_{t_2}^{t_2+t} \boldsymbol{\Phi}_{t_1}^{t_2} = \boldsymbol{\Phi}_{t_1}^{t_2+t}$, we have,
\begin{gather}
    \label{eq:sim1}
        \mathcal{P}_{t_2}^{t_2+T_y} \boldsymbol{\Phi}_{t_1}^{t_2} = \boldsymbol{\Phi}_{t_1}^{t_2+T_y}.
\end{gather}
Now, $\boldsymbol{\Phi}_{t_1}^{t_2+T_y} = \boldsymbol{\Phi}_{t_1+T_y}^{t_2+T_y} \boldsymbol{\Phi}_{t_1}^{t_1+T_y}$. Since $\boldsymbol{A}(t+T_y) = \boldsymbol A(t)$, we have:
\begin{gather}
    \label{eq:sim2}
    \boldsymbol{\Phi}_{t_1}^{t_2} = \boldsymbol{\Phi}_{t_1+T_y}^{t_2+T_y} \implies 
    \boldsymbol{\Phi}_{t_1}^{t_2+T_y} = \boldsymbol{\Phi}_{t_1}^{t_2} \boldsymbol{\Phi}_{t_1}^{t_1+T_y}.
\end{gather}
Using this in equation \eqref{eq:sim1} we get,
\begin{gather}
    \label{eq:sim3}
    \mathcal{P}_{t_2}^{t_2+T_y} \boldsymbol{\Phi}_{t_1}^{t_2} = \boldsymbol{\Phi}_{t_1}^{t_2}\mathcal{P}_{t_1}^{t_1+T_y}.
\end{gather}
Thus, the two Poincare maps are related by a similarity transformation. The eigenvalues of these maps are therefore the same, and given by $1,\, e^{\pm i \theta}$, where $\theta$ is only a function of $T_y$ and hence $\theta = \theta(H)$.

\section{Parametric equation of a set of bodies with only two planes of symmetry}

A class of bodies ranging from settlers through drifters to flutterers can be constructed using a parameterization as described here. The centerline of the bodies is a bent filament which is parameterized by a parameter $s$ as: 
\begin{gather}
    \label{eq:filament}
    \boldsymbol{r}_c(s) = a\cos(k s)\boldsymbol{p}_2 + s \boldsymbol{p}_3,\, s\in [-1,1],    
\end{gather}
where the parameterization is in non-dimensional form with half the size of the bent filament as the length scale, $k=\pi/2$ and $a$ is the dimensionless bending parameter (see red filament in figure (9) in the main paper). To construct a 2D surface, cross sections are drawn around this filament ranging from circular near the edges ($s=\pm 1$) to concavo-convex near the centre $s=0$. The normal vector $\boldsymbol{n}_c$ and binormal vector $\boldsymbol{b}_c$ to the bent filament are given by
\begin{subequations}
    \label{eq:nHatbNhat}
    \begin{gather}
        \boldsymbol{n}_c(s) = \frac{-\boldsymbol{p}_2 - k a \sin(k s) \boldsymbol{p}_3}{\sqrt{1+k^2 a^2 \sin(k s)^2}} \\
        \boldsymbol{b}_c(s) = \boldsymbol{p}_1. 
    \end{gather}
\end{subequations}
The cross-section of the body uses another parameter $\phi$ and can be constructed in the plane spanned by $\boldsymbol{n}_c$ and $\boldsymbol{b}_c$. The parametric equation of the set of bodies is given by
\begin{gather}
    \label{eq:paramEqn}
    \boldsymbol{r}(s,\, \phi) = \boldsymbol{r}_c(s) + R(s)\{ [(1-\hat b(s) )\cos(\phi) \nonumber\\
    + \hat b(s)\cos(2\phi)]\boldsymbol{n}_c(s)  + [(1 + \hat b(s)) \sin(\phi) \nonumber\\
    + \hat b(s)^3 \sin(2 \phi)]\boldsymbol{b}_c(s) \},
\end{gather}
where $R(s) = R_0\sqrt{1-s^4}$ describes the dimensionless size of the cross-section and $\hat{b}(s) \equiv b(1-s^2)$ describes the concavity of the cross-section (see figure (9) in the main paper). The value of $b=0$ corresponds to a circular cross-section with the radius of the circle given by $R_0$ at the centre of the filament $s=0$. This corresponds to a bent capsule, which is a settler. As $b$ is varied, the cross-section becomes more and more concavo-convex. 
For a given $b>0.2$, increasing the bending parameter $a$ takes us from a settler to a flutterer, going through a drifter shape for a particular value of $a$. Further increase in $a$, however, results in the flutterer going back to a settler. 
The value of $R_0$ is kept fixed to be $0.3$ for the numerical evaluation of the resistance matrix $\mathcal{R}$. 

\section{Mobility matrix for di-bilaterals}
\label{sec:CTensor}
Since the mobility problem (which involves solving for the dynamics of a body given force and torque) is being solved, we show the mobility matrix below, for completeness. The mobility matrix $\mathcal{M}$ is simply the inverse of the resistance matrix $\mathcal{R}$ and is given by:
\begin{gather}
\label{eq:Mmat}
    \mathcal{M} = 
    \begin{pmatrix}
        G_1 & 0 & 0 & 0 & 0 & -\alpha_r\\
        0 & G_2 & 0 & 0 & 0 & 0 \\
        0 & 0 & G_3 & -\alpha_p & 0 & 0 \\
        0 & 0 & -\alpha_p & A_3 G_3 C_1^{-1}  & 0 & 0 \\
        0 & 0 & 0 & 0 & C_2^{-1} & 0 \\
        -\alpha_r & 0 & 0 & 0 & 0 & A_1G_1 C_3^{-1} \\
    \end{pmatrix}.
\end{gather}
The coupling tensor $\mathcal{C}$ which characterizes the translation-rotation coupling is defined for any sedimenting body as
\begin{equation}
    \label{eq:couplingTensor}    
    \boldsymbol\Omega=\mathcal{C}\cdot\boldsymbol{F}, 
\end{equation}
where $\boldsymbol\Omega$ is the angular velocity of the body and $\boldsymbol{F}$ is the buoyancy corrected weight of the body. 
Note that the mobility coupling tensor $\mathcal{C}$ is the off-diagonal part of the full mobility matrix and in the $\{\boldsymbol{p}_i\}_{i \in \{1,2,3 \}}$ basis is given by:
\begin{gather}
    \label{eq:Tmat}
    \mathcal{C} \equiv 
    \begin{pmatrix}
        0 & 0 & -\alpha_p \\
        0 & 0 & 0 \\
        -\alpha_r & 0 & 0
    \end{pmatrix}.
\end{gather}

\section{Robustness of centre fixed points upon shifting mobility centre from centre of mass}

Equation \eqref{eq:couplingTensor} defines the coupling tensor $\mathcal{C}$ for any arbitrary-shaped body.
It has been reported \cite{witten2020review, moths2013orientational} that any body for which $\mathcal{C}$ is symmetric has six fixed points in the $\boldsymbol{p}_y$ dynamics. Two of these are saddles and four are centre fixed points. Typically, the centre fixed points undergo bifurcations and become stable or unstable points when the coupling tensor $\mathcal{C}$ departs from being symmetric\footnote{This happens when the mobility centre of a body does not coincide with its centre of mass.}. For flutterers, however, we show that the stability properties of the centre fixed points do not change even when the coupling tensor $\mathcal{C}$ is asymmetric. The reason for this is that the condition for the change in stability, as given in \cite{GONZALEZ_GRAF_MADDOCKS_2004}, is not met. This condition is stated as follows: The centre fixed points undergo bifurcations if
$\mathcal{C}$ has three distinct real eigenvalues, and the corresponding normalized eigenvectors $\boldsymbol\eta_k$ satisfy 
$$
\boldsymbol\eta_i \cdot\boldsymbol\eta_j \neq (\boldsymbol\eta_i\cdot \boldsymbol\eta_k)(\boldsymbol\eta_j\cdot \boldsymbol\eta_k)\quad (i,\, j,\, k\quad \text{distinct}),
$$
for values of the index $k$ corresponding to the maximum and minimum eigenvalues.
For di-bilaterals, the eigen-system of $\mathcal{C}$ (see equation \eqref{eq:Tmat}) is given by $\mathcal{C}\cdot\boldsymbol{\eta}_i = \lambda_i \boldsymbol\eta_i,\, i\in \{1,2,3 \}$, where:
\begin{gather*}    
\lambda_1 = 0,\quad  \boldsymbol\eta_1 =  (0,1,0), \\
\lambda_2 = -\sqrt{\alpha_p \alpha_r},\quad \boldsymbol\eta_2 = \left(-\sqrt{\frac{\alpha_p}{\alpha_p+\alpha_r}}, 0, \sqrt{\frac{\alpha_r}{\alpha_p+\alpha_r}}\right), \\
\lambda_3=\sqrt{\alpha_p \alpha_r},\quad \boldsymbol\eta_3 = \left(\sqrt{\frac{\alpha_p}{\alpha_p+\alpha_r}}, 0, \sqrt{\frac{\alpha_r}{\alpha_p+\alpha_r}}\right).
\end{gather*}
Therefore,
\begin{equation}
\boldsymbol\eta_1\cdot \boldsymbol\eta_2 = 0 = (\boldsymbol\eta_1\cdot \boldsymbol\eta_3) (\boldsymbol\eta_2\cdot \boldsymbol\eta_3),    
\end{equation}
where $\lambda_3$ is the maximum eigenvalue of $\mathcal{C}$.
Thus, flutterers do not satisfy the condition in \cite{GONZALEZ_GRAF_MADDOCKS_2004}, and there is no bifurcation in their centre fixed points. Further, we have shown that settlers have only two fixed points, not six. They do not satisfy the requirement of three distinct real eigenvalues for $\mathcal{C}$. Drifters are a special case where all eigenvalues are $0$.

\section{Dependence of settling velocities on the conserved quantity}

\begin{figure}[h]
\centering
\includegraphics[scale=0.21]{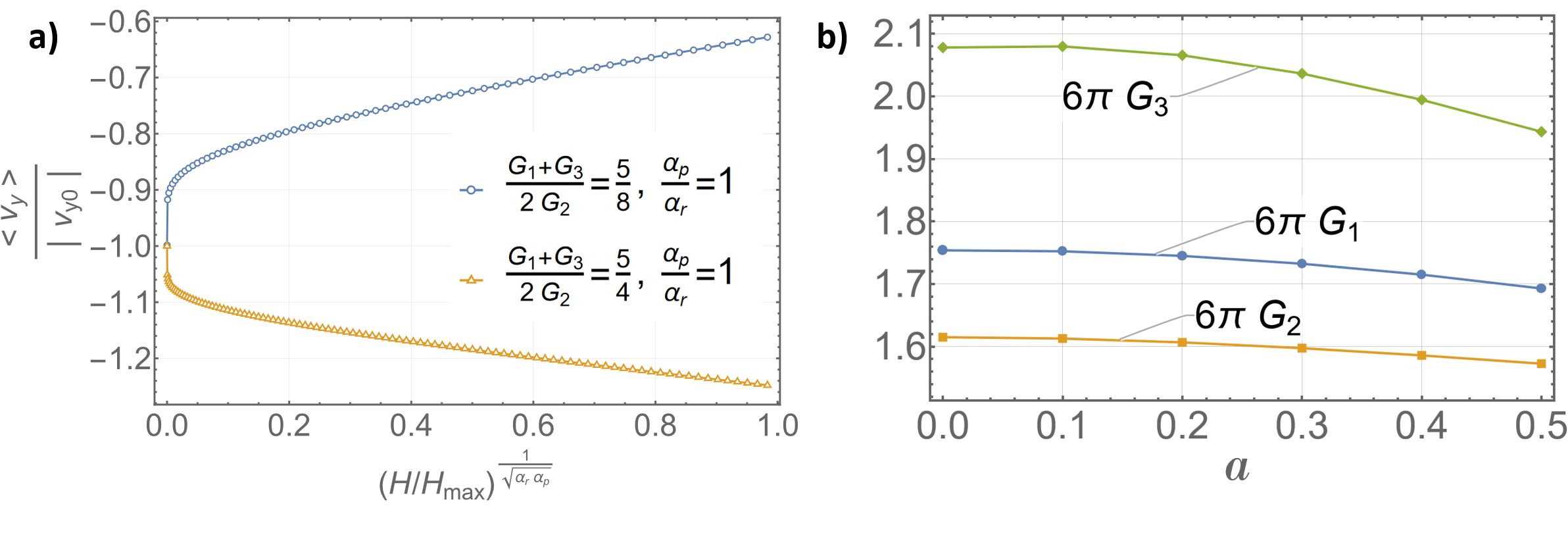}
\caption{\label{fig:meanVbyHmax} a) For a given body, the mean vertical velocity is a function only of $H$, and two samples are shown here.
b) Dependence of parameters determining the settling velocity on the bending parameter $a$, for a fixed $b$ = 0.4.
}
\end{figure}

The translational velocity of 
di-bilaterals can be obtained from equations (1) and (2) of the main paper. In the lab frame 
\begin{subequations}
    \label{eq:Veqn}
    \begin{gather}
        v_{x} = -G_1 p_{1x}p_{1y} -G_2 p_{2x}p_{2y} -G_3 p_{3x}p_{3y}, \\
        v_{y} = -G_1 p_{1y}^2 -G_2 p_{2y}^2 -G_3 p_{3y}^2, \\
        v_{z} = -G_1 p_{1z}p_{1y} -G_2 p_{2z}p_{2y} -G_3 p_{3z}p_{3y},     
    \end{gather}
\end{subequations}
where $(p_{1z}, p_{2z}, p_{3z}) \equiv \boldsymbol{p}_z \equiv \boldsymbol{p}_x\times \boldsymbol{p}_y$, $G_2 \equiv 1/A_2$ and
\begin{subequations}
\label{eq:Gparams}
\begin{gather}
    G_1 \equiv \frac{C_3}{A_1C_3-B_{31}^2},\quad  G_3 \equiv \frac{C_1}{A_3C_1-B_{13}^2}.
\end{gather}    
\end{subequations}
Viscous dissipation requires $G_1, G_2, G_3>0$ \cite{happel2012low, kim2013microhydrodynamics, guazzelli2011physical, graham2018microhydrodynamics}. Equation \eqref{eq:Veqn} shows the direct dependence of the translational velocity on the orientation of the body. 
 For $H=0$, the settling velocity $v_{y}$ reaches $v_{y0}=-G_2$ exponentially fast in time with the rate $|\alpha_p|$ or $|\alpha_r|$, depending on the initial condition, and for $H=H_{\text{max}}$, $v_{y}=v_{ym}\equiv-(G_1\alpha_p+G_3\alpha_r)/(\alpha_p+\alpha_r)$. The shape of the body determines whether $\langle v_y\rangle$ increases or decreases with $H$, and a sample of each is shown in figure \ref{fig:meanVbyHmax}. 

\section{Comment on the sedimentation dynamics of generic shape}
\subsection{Angular dynamics}

The rate of change of body attached coordinate axes are given by:
\begin{equation}
    \label{eq:piDot}
    \Dot{\boldsymbol{p}_i} = \boldsymbol\Omega\times\boldsymbol{p}_i = (\mathcal{C}_g\cdot\boldsymbol{F})\times \boldsymbol{p}_i, \quad i \in \{1,2,3 \},
\end{equation}
where $\mathcal{C}_g$ is the coupling tensor of a generic body (defined in section \ref{sec:CTensor}, equation \eqref{eq:couplingTensor}) and $\boldsymbol{F}$ is the dimensionless buoyancy corrected weight of the body. 
For generic bodies $\mathcal{C}$ has a more fuller form than for the di-bilaterials. 
In the $\{\boldsymbol{p}_i\}_{i \in \{1,2,3 \}}$ basis $\mathcal{C}_g$ is a shape dependent constant matrix and the dimensionless force is $\boldsymbol{F}=-\boldsymbol{\hat{y}}=-(p_{1y} \boldsymbol{p}_1 + p_{2y} \boldsymbol{p}_2 + p_{3y} \boldsymbol{p}_3)$\footnote{Note that gravity is taken to be along $-\boldsymbol{\hat{y}}$}. Therefore, from equation $\eqref{eq:piDot}$:
\begin{equation}
    \label{eq:pyGen1}
    \Dot{\boldsymbol{p}_i}\cdot\boldsymbol{\hat{y}} = \Dot{p_{iy}} = \boldsymbol{\hat{y}}\cdot[(\mathcal{C}_g\cdot\boldsymbol{F})\times\boldsymbol{p}_i] = \boldsymbol{p}_i\cdot[\boldsymbol{\hat{y}}\times(\mathcal{C}_g\cdot\boldsymbol{F})]
\end{equation}
In the $\{\boldsymbol{p}_i\}_{i \in \{1,2,3 \}}$ basis, $\boldsymbol{\hat{y}} = (p_{1y},p_{2y},p_{3y} ) = -\boldsymbol{F}$. Using this in equation \eqref{eq:pyGen1} we have:
\begin{equation}
    \label{eq:pyGen2}
    \Dot{p_{iy}} = [\boldsymbol{\hat{y}}\times(\mathcal{C}_g\cdot\boldsymbol{F})]_i = -[(p_{1y},p_{2y}, p_{3y} )\times\mathcal{C}_g\cdot(p_{1y},p_{2y}, p_{3y} )]_i
\end{equation}
Therefore, the dynamics of the vertical projections $\boldsymbol{p}_y$ can be written in a compact form as:
\begin{equation}
    \label{eq:pyGeneral}
    \Dot{\boldsymbol{p}_y} = (\mathcal{C}_g\cdot \boldsymbol{p}_y)\times \boldsymbol{p}_y.
\end{equation}
This shows explicitly that even for generic bodies the $\boldsymbol{p}_y$ dynamics is closed in itself, i.e. it doesn't depend on the $\boldsymbol{p}_x$. This can also be seen from the rotational invariance in the body frame about the gravity axis. This compact formulation can also be found in \cite{witten2020review, moths2013orientational}.

The previous studies on sedimentation of a generic body \cite{witten2020review, moths2013orientational, GONZALEZ_GRAF_MADDOCKS_2004, makino2005sedimentation} have shown existence of the stable, unstable as well as centre fixed points in the $\boldsymbol{p}_y$ dynamics. The most interesting $\boldsymbol{p}_y$ dynamics reported by \cite{makino2005sedimentation} and \cite{moths2013orientational} are the closed periodic orbits and limit cycles in $\boldsymbol{p}_y$ dynamics. Thus, asymptotically, $\boldsymbol{p}_y$ dynamics can be atmost periodic, if not steady. 

The remaining part of the angular dynamics involves analyzing the $\boldsymbol{p}_x$ dynamics, which, to the best of our knowledge, has not been done before. Similar to the $\boldsymbol{p}_y$ dynamical system the $\boldsymbol{p}_x$ dynamics can be written compactly as:
\begin{gather}
    \label{eq:pxEqn}
    \Dot{\boldsymbol{p}_x} = (\mathcal{C}_g\cdot\boldsymbol{p_y}) \times \boldsymbol{p}_x \equiv \boldsymbol{A}[\boldsymbol{p}_y(t)] \boldsymbol{p}_x,
\end{gather}
where $\boldsymbol{A}_{ik} = -\epsilon_{ijk} (\mathcal{C}_g\cdot\boldsymbol{p}_y)_{j}$ and $\mathcal{C}_g$ is the coupling matrix for a generic body written in the body frame (which will have a fuller form unlike the one given by eq. \eqref{eq:Tmat} for di-bilaterals). This shows that the $\boldsymbol{p}_x$ is still a linear non-autonomous dynamical system driven by $\boldsymbol{p}_y$ dynamics. If $\boldsymbol{p}_y$ settles into a stable fixed point the $\boldsymbol{p}_x$ dynamics will be in general periodic, provided $\mathcal{C}_g\cdot\boldsymbol{p}_y \neq 0$. If $\boldsymbol{p}_y$ performs a periodic orbit asymptotically, the corresponding $\boldsymbol{p}_x$ dynamics can be either quasi-periodic or periodic, which can again be studied using Floquet theory. 

\subsection{Persistent drift in a generic body}
\label{sec:PD}
The proof provided in section \ref{sec:PDFlutterer} for a di-bilateral flutterer is extended in this section to bodies of arbitrary shape which can flutter. The translational velocity of a generic body in the lab frame is given by:
\begin{gather}
    \label{eq:Vgeneric}
    v_x = \sum_{ij} c_{ij}^{(x)} p_{iy} p_{jx} \\
    v_y = \sum_{ij} c_{ij}^{(y)} p_{iy} p_{jx} \\
    v_z = \sum_{ij} c_{ij}^{(z)} p_{iy} p_{jx}, 
\end{gather}
where $c_{ij}^{(x/y/z)}$ are related to the mobility matrix of the generic body under consideration.

The case where where $\boldsymbol{p}_y$ reaches a stable fixed point has been discussed in \cite{witten2020review}. For completeness, we state their results below. As discussed in \cite{witten2020review}, there are the following two possibilities when $\boldsymbol{p}_y$ is at a fixed point:
\begin{enumerate}
    \item $\mathcal{C}_g \cdot \boldsymbol{p}_y^* = \boldsymbol{0}$: In this case the body ceases to rotate (just like settlers and drifters for di-bilaterals). The $\boldsymbol{p}_x$ becomes a constant and $v_x$, $v_z$ or both can be non-zero, resulting in persistent drift.
    \item $\mathcal{C}_g\cdot \boldsymbol{p}_y^* = \lambda \boldsymbol{p}_y^*, \, \lambda\neq 0$: In this case the $\boldsymbol{p}_x$ performs a periodic orbit, resulting in no persistent drift in one time period. Note that $\Dot{\boldsymbol{p}_x} = \lambda \boldsymbol{p}_y^*\times \boldsymbol{p}_x \neq 0$ and hence $\boldsymbol{p}_x$ cannot be static. 
    In this case, the angular velocity of the body $\boldsymbol{\Omega}$ align with the gravity and the body falls in a helical trajectory. 
\end{enumerate}

It is significant to note  that the $\boldsymbol{p}_y$ dynamical system for an arbitrary-shaped body is still quadratic in $\boldsymbol{p}_y$ and the $\boldsymbol{p}_x$ dynamical system has the same form as that for di-bilaterals.
For the case where $\boldsymbol{p}_y$ remains in a closed periodic orbit, the proof given in section \ref{sec:PDFlutterer} works for an arbitrary body as well, and so there is a persistent drift only when the time period of $\boldsymbol{p}_y$ dynamics is exactly the same as the time period of the $\boldsymbol{p}_x$ dynamics. Since the limiting behaviours of the $\boldsymbol{p}_y$ dynamics are either fixed points or periodic orbits \cite{makino2005sedimentation, moths2013orientational, witten2020review}, the cases discussed are exhaustive. Therefore, we conclude that \textit{any arbitrary body that performs periodic/quasi-periodic motion cannot drift persistently, except for the special case where its frequency of the $\boldsymbol{p}_x$ dynamics is exactly equal to the frequency
of the $\boldsymbol{p}_y$ dynamics.}

\section{Description of videos}

The three videos show characteristic rotational dynamics of each of the three classes of bodies discussed in the main paper. The black coordinate axes denote the lab-fixed coordinate system and blue ones denote the body-fixed coordinate system. For clarity, the vertical motion of the bodies is not shown in the videos. Gravity is along $-y$ axis. The black curve in the yellow plane shows the horizontal displacement of the body as it falls under gravity.
\begin{itemize}
    \item[] \textbf{Video 1: Settler} $-$ A settler eventually falls with one of its principal axes along gravity. The horizontal drift goes to zero as it reaches its stable orientation.
    \item[] \textbf{Video 2: Drifter} $-$ A drifter eventually falls with its principal axes making certain angles with respect to gravity. There is a persistent horizontal drift in this case. 
    \item[] \textbf{Video 3: Flutterer} $-$ A flutterer keeps on rotating as it falls, executing intricate patterns in the horizontal plane. However, there is no persistent horizontal drift.
\end{itemize}



\nocite{*}
\bibliography{ref}

\providecommand{\noopsort}[1]{}\providecommand{\singleletter}[1]{#1}%
\begin{thebibliography}{31}%
\makeatletter
\providecommand \@ifxundefined [1]{%
 \@ifx{#1\undefined}
}%
\providecommand \@ifnum [1]{%
 \ifnum #1\expandafter \@firstoftwo
 \else \expandafter \@secondoftwo
 \fi
}%
\providecommand \@ifx [1]{%
 \ifx #1\expandafter \@firstoftwo
 \else \expandafter \@secondoftwo
 \fi
}%
\providecommand \natexlab [1]{#1}%
\providecommand \enquote  [1]{``#1''}%
\providecommand \bibnamefont  [1]{#1}%
\providecommand \bibfnamefont [1]{#1}%
\providecommand \citenamefont [1]{#1}%
\providecommand \href@noop [0]{\@secondoftwo}%
\providecommand \href [0]{\begingroup \@sanitize@url \@href}%
\providecommand \@href[1]{\@@startlink{#1}\@@href}%
\providecommand \@@href[1]{\endgroup#1\@@endlink}%
\providecommand \@sanitize@url [0]{\catcode `\\12\catcode `\$12\catcode `\&12\catcode `\#12\catcode `\^12\catcode `\_12\catcode `\%12\relax}%
\providecommand \@@startlink[1]{}%
\providecommand \@@endlink[0]{}%
\providecommand \url  [0]{\begingroup\@sanitize@url \@url }%
\providecommand \@url [1]{\endgroup\@href {#1}{\urlprefix }}%
\providecommand \urlprefix  [0]{URL }%
\providecommand \Eprint [0]{\href }%
\providecommand \doibase [0]{https://doi.org/}%
\providecommand \selectlanguage [0]{\@gobble}%
\providecommand \bibinfo  [0]{\@secondoftwo}%
\providecommand \bibfield  [0]{\@secondoftwo}%
\providecommand \translation [1]{[#1]}%
\providecommand \BibitemOpen [0]{}%
\providecommand \bibitemStop [0]{}%
\providecommand \bibitemNoStop [0]{.\EOS\space}%
\providecommand \EOS [0]{\spacefactor3000\relax}%
\providecommand \BibitemShut  [1]{\csname bibitem#1\endcsname}%
\let\auto@bib@innerbib\@empty
\bibitem [{\citenamefont {Field}\ \emph {et~al.}(1997)\citenamefont {Field}, \citenamefont {Klaus}, \citenamefont {Moore},\ and\ \citenamefont {Nori}}]{field1997chaotic}%
  \BibitemOpen
  \bibfield  {author} {\bibinfo {author} {\bibfnamefont {S.~B.}\ \bibnamefont {Field}}, \bibinfo {author} {\bibfnamefont {M.}~\bibnamefont {Klaus}}, \bibinfo {author} {\bibfnamefont {M.}~\bibnamefont {Moore}},\ and\ \bibinfo {author} {\bibfnamefont {F.}~\bibnamefont {Nori}},\ }\bibfield  {title} {\bibinfo {title} {Chaotic dynamics of falling disks},\ }\href@noop {} {\bibfield  {journal} {\bibinfo  {journal} {Nature}\ }\textbf {\bibinfo {volume} {388}},\ \bibinfo {pages} {252} (\bibinfo {year} {1997})}\BibitemShut {NoStop}%
\bibitem [{\citenamefont {Zhong}\ \emph {et~al.}(2011)\citenamefont {Zhong}, \citenamefont {Chen},\ and\ \citenamefont {Lee}}]{zhong2011experimental}%
  \BibitemOpen
  \bibfield  {author} {\bibinfo {author} {\bibfnamefont {H.}~\bibnamefont {Zhong}}, \bibinfo {author} {\bibfnamefont {S.}~\bibnamefont {Chen}},\ and\ \bibinfo {author} {\bibfnamefont {C.}~\bibnamefont {Lee}},\ }\bibfield  {title} {\bibinfo {title} {Experimental study of freely falling thin disks: Transition from planar zigzag to spiral},\ }\href@noop {} {\bibfield  {journal} {\bibinfo  {journal} {Physics of Fluids}\ }\textbf {\bibinfo {volume} {23}} (\bibinfo {year} {2011})}\BibitemShut {NoStop}%
\bibitem [{\citenamefont {Auguste}\ \emph {et~al.}(2013)\citenamefont {Auguste}, \citenamefont {Magnaudet},\ and\ \citenamefont {Fabre}}]{auguste2013falling}%
  \BibitemOpen
  \bibfield  {author} {\bibinfo {author} {\bibfnamefont {F.}~\bibnamefont {Auguste}}, \bibinfo {author} {\bibfnamefont {J.}~\bibnamefont {Magnaudet}},\ and\ \bibinfo {author} {\bibfnamefont {D.}~\bibnamefont {Fabre}},\ }\bibfield  {title} {\bibinfo {title} {Falling styles of disks},\ }\href@noop {} {\bibfield  {journal} {\bibinfo  {journal} {Journal of Fluid Mechanics}\ }\textbf {\bibinfo {volume} {719}},\ \bibinfo {pages} {388} (\bibinfo {year} {2013})}\BibitemShut {NoStop}%
\bibitem [{\citenamefont {Happel}\ and\ \citenamefont {Brenner}(2012)}]{happel2012low}%
  \BibitemOpen
  \bibfield  {author} {\bibinfo {author} {\bibfnamefont {J.}~\bibnamefont {Happel}}\ and\ \bibinfo {author} {\bibfnamefont {H.}~\bibnamefont {Brenner}},\ }\href@noop {} {\emph {\bibinfo {title} {Low Reynolds number hydrodynamics: with special applications to particulate media}}},\ Vol.~\bibinfo {volume} {1}\ (\bibinfo  {publisher} {Springer Science \& Business Media},\ \bibinfo {year} {2012})\BibitemShut {NoStop}%
\bibitem [{\citenamefont {Kim}\ and\ \citenamefont {Karrila}(2013)}]{kim2013microhydrodynamics}%
  \BibitemOpen
  \bibfield  {author} {\bibinfo {author} {\bibfnamefont {S.}~\bibnamefont {Kim}}\ and\ \bibinfo {author} {\bibfnamefont {S.~J.}\ \bibnamefont {Karrila}},\ }\href@noop {} {\emph {\bibinfo {title} {Microhydrodynamics: principles and selected applications}}}\ (\bibinfo  {publisher} {Butterworth-Heinemann},\ \bibinfo {year} {2013})\BibitemShut {NoStop}%
\bibitem [{\citenamefont {Witten}\ and\ \citenamefont {Diamant}(2020)}]{witten2020review}%
  \BibitemOpen
  \bibfield  {author} {\bibinfo {author} {\bibfnamefont {T.~A.}\ \bibnamefont {Witten}}\ and\ \bibinfo {author} {\bibfnamefont {H.}~\bibnamefont {Diamant}},\ }\bibfield  {title} {\bibinfo {title} {A review of shaped colloidal particles in fluids: anisotropy and chirality},\ }\href@noop {} {\bibfield  {journal} {\bibinfo  {journal} {Reports on progress in physics}\ }\textbf {\bibinfo {volume} {83}},\ \bibinfo {pages} {116601} (\bibinfo {year} {2020})}\BibitemShut {NoStop}%
\bibitem [{\citenamefont {Guazzelli}\ and\ \citenamefont {Morris}(2011)}]{guazzelli2011physical}%
  \BibitemOpen
  \bibfield  {author} {\bibinfo {author} {\bibfnamefont {E.}~\bibnamefont {Guazzelli}}\ and\ \bibinfo {author} {\bibfnamefont {J.~F.}\ \bibnamefont {Morris}},\ }\href@noop {} {\emph {\bibinfo {title} {A physical introduction to suspension dynamics}}},\ Vol.~\bibinfo {volume} {45}\ (\bibinfo  {publisher} {Cambridge University Press},\ \bibinfo {year} {2011})\BibitemShut {NoStop}%
\bibitem [{\citenamefont {Graham}(2018)}]{graham2018microhydrodynamics}%
  \BibitemOpen
  \bibfield  {author} {\bibinfo {author} {\bibfnamefont {M.~D.}\ \bibnamefont {Graham}},\ }\href@noop {} {\emph {\bibinfo {title} {Microhydrodynamics, Brownian motion, and complex fluids}}},\ Vol.~\bibinfo {volume} {58}\ (\bibinfo  {publisher} {Cambridge University Press},\ \bibinfo {year} {2018})\BibitemShut {NoStop}%
\bibitem [{\citenamefont {Palusa}\ \emph {et~al.}(2018)\citenamefont {Palusa}, \citenamefont {de~Graaf}, \citenamefont {Brown},\ and\ \citenamefont {Morozov}}]{PhysRevFluids.3.124301}%
  \BibitemOpen
  \bibfield  {author} {\bibinfo {author} {\bibfnamefont {M.}~\bibnamefont {Palusa}}, \bibinfo {author} {\bibfnamefont {J.}~\bibnamefont {de~Graaf}}, \bibinfo {author} {\bibfnamefont {A.}~\bibnamefont {Brown}},\ and\ \bibinfo {author} {\bibfnamefont {A.}~\bibnamefont {Morozov}},\ }\bibfield  {title} {\bibinfo {title} {Sedimentation of a rigid helix in viscous media},\ }\href {https://doi.org/10.1103/PhysRevFluids.3.124301} {\bibfield  {journal} {\bibinfo  {journal} {Phys. Rev. Fluids}\ }\textbf {\bibinfo {volume} {3}},\ \bibinfo {pages} {124301} (\bibinfo {year} {2018})}\BibitemShut {NoStop}%
\bibitem [{\citenamefont {Krapf}\ \emph {et~al.}(2009)\citenamefont {Krapf}, \citenamefont {Witten},\ and\ \citenamefont {Keim}}]{krapf2009chiral}%
  \BibitemOpen
  \bibfield  {author} {\bibinfo {author} {\bibfnamefont {N.~W.}\ \bibnamefont {Krapf}}, \bibinfo {author} {\bibfnamefont {T.~A.}\ \bibnamefont {Witten}},\ and\ \bibinfo {author} {\bibfnamefont {N.~C.}\ \bibnamefont {Keim}},\ }\bibfield  {title} {\bibinfo {title} {Chiral sedimentation of extended objects in viscous media},\ }\href@noop {} {\bibfield  {journal} {\bibinfo  {journal} {Physical Review E}\ }\textbf {\bibinfo {volume} {79}},\ \bibinfo {pages} {056307} (\bibinfo {year} {2009})}\BibitemShut {NoStop}%
\bibitem [{\citenamefont {Collins}\ \emph {et~al.}(2021)\citenamefont {Collins}, \citenamefont {Hamati}, \citenamefont {Candelier}, \citenamefont {Gustavsson}, \citenamefont {Mehlig},\ and\ \citenamefont {Voth}}]{collins2021lord}%
  \BibitemOpen
  \bibfield  {author} {\bibinfo {author} {\bibfnamefont {D.}~\bibnamefont {Collins}}, \bibinfo {author} {\bibfnamefont {R.~J.}\ \bibnamefont {Hamati}}, \bibinfo {author} {\bibfnamefont {F.}~\bibnamefont {Candelier}}, \bibinfo {author} {\bibfnamefont {K.}~\bibnamefont {Gustavsson}}, \bibinfo {author} {\bibfnamefont {B.}~\bibnamefont {Mehlig}},\ and\ \bibinfo {author} {\bibfnamefont {G.~A.}\ \bibnamefont {Voth}},\ }\bibfield  {title} {\bibinfo {title} {Lord kelvin's isotropic helicoid},\ }\href@noop {} {\bibfield  {journal} {\bibinfo  {journal} {Physical Review Fluids}\ }\textbf {\bibinfo {volume} {6}},\ \bibinfo {pages} {074302} (\bibinfo {year} {2021})}\BibitemShut {NoStop}%
\bibitem [{\citenamefont {Brenner}(1964)}]{brenner1964stokes}%
  \BibitemOpen
  \bibfield  {author} {\bibinfo {author} {\bibfnamefont {H.}~\bibnamefont {Brenner}},\ }\bibfield  {title} {\bibinfo {title} {The stokes resistance of an arbitrary particle—ii: An extension},\ }\href@noop {} {\bibfield  {journal} {\bibinfo  {journal} {Chemical Engineering Science}\ }\textbf {\bibinfo {volume} {19}},\ \bibinfo {pages} {599} (\bibinfo {year} {1964})}\BibitemShut {NoStop}%
\bibitem [{\citenamefont {Miara}\ \emph {et~al.}(2024)\citenamefont {Miara}, \citenamefont {Vaquero-Stainer}, \citenamefont {Pihler-Puzovi{\'c}}, \citenamefont {Heil},\ and\ \citenamefont {Juel}}]{miara2024dynamics}%
  \BibitemOpen
  \bibfield  {author} {\bibinfo {author} {\bibfnamefont {T.}~\bibnamefont {Miara}}, \bibinfo {author} {\bibfnamefont {C.}~\bibnamefont {Vaquero-Stainer}}, \bibinfo {author} {\bibfnamefont {D.}~\bibnamefont {Pihler-Puzovi{\'c}}}, \bibinfo {author} {\bibfnamefont {M.}~\bibnamefont {Heil}},\ and\ \bibinfo {author} {\bibfnamefont {A.}~\bibnamefont {Juel}},\ }\bibfield  {title} {\bibinfo {title} {Dynamics of inertialess sedimentation of a rigid u-shaped disk},\ }\href@noop {} {\bibfield  {journal} {\bibinfo  {journal} {Communications Physics}\ }\textbf {\bibinfo {volume} {7}},\ \bibinfo {pages} {47} (\bibinfo {year} {2024})}\BibitemShut {NoStop}%
\bibitem [{\citenamefont {Cosentino~Lagomarsino}\ \emph {et~al.}(2005)\citenamefont {Cosentino~Lagomarsino}, \citenamefont {Pagonabarraga},\ and\ \citenamefont {Lowe}}]{PhysRevLett.94.148104}%
  \BibitemOpen
  \bibfield  {author} {\bibinfo {author} {\bibfnamefont {M.}~\bibnamefont {Cosentino~Lagomarsino}}, \bibinfo {author} {\bibfnamefont {I.}~\bibnamefont {Pagonabarraga}},\ and\ \bibinfo {author} {\bibfnamefont {C.~P.}\ \bibnamefont {Lowe}},\ }\bibfield  {title} {\bibinfo {title} {Hydrodynamic induced deformation and orientation of a microscopic elastic filament},\ }\href {https://doi.org/10.1103/PhysRevLett.94.148104} {\bibfield  {journal} {\bibinfo  {journal} {Phys. Rev. Lett.}\ }\textbf {\bibinfo {volume} {94}},\ \bibinfo {pages} {148104} (\bibinfo {year} {2005})}\BibitemShut {NoStop}%
\bibitem [{\citenamefont {Thorp}\ and\ \citenamefont {Lister}(2019)}]{thorp2019motion}%
  \BibitemOpen
  \bibfield  {author} {\bibinfo {author} {\bibfnamefont {I.~R.}\ \bibnamefont {Thorp}}\ and\ \bibinfo {author} {\bibfnamefont {J.~R.}\ \bibnamefont {Lister}},\ }\bibfield  {title} {\bibinfo {title} {Motion of a non-axisymmetric particle in viscous shear flow},\ }\href@noop {} {\bibfield  {journal} {\bibinfo  {journal} {Journal of Fluid Mechanics}\ }\textbf {\bibinfo {volume} {872}},\ \bibinfo {pages} {532} (\bibinfo {year} {2019})}\BibitemShut {NoStop}%
\bibitem [{\citenamefont {Makino}\ \emph {et~al.}(2005)\citenamefont {Makino} \emph {et~al.}}]{makino2005sedimentation}%
  \BibitemOpen
  \bibfield  {author} {\bibinfo {author} {\bibfnamefont {M.}~\bibnamefont {Makino}} \emph {et~al.},\ }\bibfield  {title} {\bibinfo {title} {Sedimentation of particles of general shape},\ }\href@noop {} {\bibfield  {journal} {\bibinfo  {journal} {Physics of Fluids}\ }\textbf {\bibinfo {volume} {17}} (\bibinfo {year} {2005})}\BibitemShut {NoStop}%
\bibitem [{\citenamefont {Moths}\ and\ \citenamefont {Witten}(2013)}]{moths2013orientational}%
  \BibitemOpen
  \bibfield  {author} {\bibinfo {author} {\bibfnamefont {B.}~\bibnamefont {Moths}}\ and\ \bibinfo {author} {\bibfnamefont {T.}~\bibnamefont {Witten}},\ }\bibfield  {title} {\bibinfo {title} {Orientational ordering of colloidal dispersions by application of time-dependent external forces},\ }\href@noop {} {\bibfield  {journal} {\bibinfo  {journal} {Physical Review E—Statistical, Nonlinear, and Soft Matter Physics}\ }\textbf {\bibinfo {volume} {88}},\ \bibinfo {pages} {022307} (\bibinfo {year} {2013})}\BibitemShut {NoStop}%
\bibitem [{sup()}]{supplemental}%
  \BibitemOpen
  \href@noop {} {}\bibinfo {note} {See Supplemental Material for additional details, which includes Ref. [29].}\BibitemShut {Stop}%
\bibitem [{\citenamefont {Tozzi}\ \emph {et~al.}(2011)\citenamefont {Tozzi}, \citenamefont {Scott}, \citenamefont {Vahey},\ and\ \citenamefont {Klingenberg}}]{tozzi2011settling}%
  \BibitemOpen
  \bibfield  {author} {\bibinfo {author} {\bibfnamefont {E.}~\bibnamefont {Tozzi}}, \bibinfo {author} {\bibfnamefont {C.}~\bibnamefont {Scott}}, \bibinfo {author} {\bibfnamefont {D.}~\bibnamefont {Vahey}},\ and\ \bibinfo {author} {\bibfnamefont {D.}~\bibnamefont {Klingenberg}},\ }\bibfield  {title} {\bibinfo {title} {Settling dynamics of asymmetric rigid fibers},\ }\href@noop {} {\bibfield  {journal} {\bibinfo  {journal} {Physics of Fluids}\ }\textbf {\bibinfo {volume} {23}} (\bibinfo {year} {2011})}\BibitemShut {NoStop}%
\bibitem [{\citenamefont {Ekiel-Je{\.z}ewska}\ and\ \citenamefont {Wajnryb}(2009)}]{ekiel2009hydrodynamic}%
  \BibitemOpen
  \bibfield  {author} {\bibinfo {author} {\bibfnamefont {M.~L.}\ \bibnamefont {Ekiel-Je{\.z}ewska}}\ and\ \bibinfo {author} {\bibfnamefont {E.}~\bibnamefont {Wajnryb}},\ }\bibfield  {title} {\bibinfo {title} {Hydrodynamic orienting of asymmetric microobjects under gravity},\ }\href@noop {} {\bibfield  {journal} {\bibinfo  {journal} {Journal of Physics: Condensed Matter}\ }\textbf {\bibinfo {volume} {21}},\ \bibinfo {pages} {204102} (\bibinfo {year} {2009})}\BibitemShut {NoStop}%
\bibitem [{\citenamefont {Ravichandran}\ and\ \citenamefont {Wettlaufer}(2023)}]{PhysRevFluids.8.L062301}%
  \BibitemOpen
  \bibfield  {author} {\bibinfo {author} {\bibfnamefont {S.}~\bibnamefont {Ravichandran}}\ and\ \bibinfo {author} {\bibfnamefont {J.~S.}\ \bibnamefont {Wettlaufer}},\ }\bibfield  {title} {\bibinfo {title} {Orientation dynamics of two-dimensional concavo-convex bodies},\ }\href {https://doi.org/10.1103/PhysRevFluids.8.L062301} {\bibfield  {journal} {\bibinfo  {journal} {Phys. Rev. Fluids}\ }\textbf {\bibinfo {volume} {8}},\ \bibinfo {pages} {L062301} (\bibinfo {year} {2023})}\BibitemShut {NoStop}%
\bibitem [{\citenamefont {Vaquero-Stainer}\ \emph {et~al.}(2024)\citenamefont {Vaquero-Stainer}, \citenamefont {Miara}, \citenamefont {Juel}, \citenamefont {Pihler-Puzovi{\'c}},\ and\ \citenamefont {Heil}}]{vaquero2024u}%
  \BibitemOpen
  \bibfield  {author} {\bibinfo {author} {\bibfnamefont {C.}~\bibnamefont {Vaquero-Stainer}}, \bibinfo {author} {\bibfnamefont {T.}~\bibnamefont {Miara}}, \bibinfo {author} {\bibfnamefont {A.}~\bibnamefont {Juel}}, \bibinfo {author} {\bibfnamefont {D.}~\bibnamefont {Pihler-Puzovi{\'c}}},\ and\ \bibinfo {author} {\bibfnamefont {M.}~\bibnamefont {Heil}},\ }\bibfield  {title} {\bibinfo {title} {U-shaped disks in stokes flow: Chiral sedimentation of a non-chiral particle},\ }\href@noop {} {\bibfield  {journal} {\bibinfo  {journal} {arXiv preprint arXiv:2406.13837}\ } (\bibinfo {year} {2024})}\BibitemShut {NoStop}%
\bibitem [{\citenamefont {GONZALEZ}\ \emph {et~al.}(2004)\citenamefont {GONZALEZ}, \citenamefont {GRAF},\ and\ \citenamefont {MADDOCKS}}]{GONZALEZ_GRAF_MADDOCKS_2004}%
  \BibitemOpen
  \bibfield  {author} {\bibinfo {author} {\bibfnamefont {O.}~\bibnamefont {GONZALEZ}}, \bibinfo {author} {\bibfnamefont {A.~B.~A.}\ \bibnamefont {GRAF}},\ and\ \bibinfo {author} {\bibfnamefont {J.~H.}\ \bibnamefont {MADDOCKS}},\ }\bibfield  {title} {\bibinfo {title} {Dynamics of a rigid body in a stokes fluid},\ }\href {https://doi.org/10.1017/S0022112004001284} {\bibfield  {journal} {\bibinfo  {journal} {Journal of Fluid Mechanics}\ }\textbf {\bibinfo {volume} {519}},\ \bibinfo {pages} {133–160} (\bibinfo {year} {2004})}\BibitemShut {NoStop}%
\bibitem [{\citenamefont {Chicone}(2006)}]{chicone2006ordinary}%
  \BibitemOpen
  \bibfield  {author} {\bibinfo {author} {\bibfnamefont {C.}~\bibnamefont {Chicone}},\ }\href@noop {} {\emph {\bibinfo {title} {Ordinary differential equations with applications}}}\ (\bibinfo  {publisher} {Springer},\ \bibinfo {year} {2006})\BibitemShut {NoStop}%
\bibitem [{\citenamefont {Bagge}\ and\ \citenamefont {Tornberg}(2021)}]{bagge2021highly}%
  \BibitemOpen
  \bibfield  {author} {\bibinfo {author} {\bibfnamefont {J.}~\bibnamefont {Bagge}}\ and\ \bibinfo {author} {\bibfnamefont {A.-K.}\ \bibnamefont {Tornberg}},\ }\bibfield  {title} {\bibinfo {title} {Highly accurate special quadrature methods for stokesian particle suspensions in confined geometries},\ }\href@noop {} {\bibfield  {journal} {\bibinfo  {journal} {International Journal for Numerical Methods in Fluids}\ }\textbf {\bibinfo {volume} {93}},\ \bibinfo {pages} {2175} (\bibinfo {year} {2021})}\BibitemShut {NoStop}%
\bibitem [{\citenamefont {Pozrikidis}(1992)}]{pozrikidis1992boundary}%
  \BibitemOpen
  \bibfield  {author} {\bibinfo {author} {\bibfnamefont {C.}~\bibnamefont {Pozrikidis}},\ }\href@noop {} {\emph {\bibinfo {title} {Boundary integral and singularity methods for linearized viscous flow}}}\ (\bibinfo  {publisher} {Cambridge university press},\ \bibinfo {year} {1992})\BibitemShut {NoStop}%
\bibitem [{\citenamefont {Pozrikidis}(2002)}]{pozrikidis2002practical}%
  \BibitemOpen
  \bibfield  {author} {\bibinfo {author} {\bibfnamefont {C.}~\bibnamefont {Pozrikidis}},\ }\href@noop {} {\emph {\bibinfo {title} {A practical guide to boundary element methods with the software library BEMLIB}}}\ (\bibinfo  {publisher} {CRC Press},\ \bibinfo {year} {2002})\BibitemShut {NoStop}%
\bibitem [{\citenamefont {Prosperetti}\ and\ \citenamefont {Tryggvason}(2009)}]{prosperetti2009computational}%
  \BibitemOpen
  \bibfield  {author} {\bibinfo {author} {\bibfnamefont {A.}~\bibnamefont {Prosperetti}}\ and\ \bibinfo {author} {\bibfnamefont {G.}~\bibnamefont {Tryggvason}},\ }\href@noop {} {\emph {\bibinfo {title} {Computational methods for multiphase flow}}}\ (\bibinfo  {publisher} {Cambridge university press},\ \bibinfo {year} {2009})\BibitemShut {NoStop}%
\bibitem [{\citenamefont {Hydon}(2000)}]{hydon2000symmetry}%
  \BibitemOpen
  \bibfield  {author} {\bibinfo {author} {\bibfnamefont {P.~E.}\ \bibnamefont {Hydon}},\ }\href@noop {} {\emph {\bibinfo {title} {Symmetry methods for differential equations: a beginner's guide}}}\ (\bibinfo  {publisher} {Cambridge University Press},\ \bibinfo {year} {2000})\ \bibinfo {note} {22}\BibitemShut {NoStop}%
\bibitem [{Note1()}]{Note1}%
  \BibitemOpen
  \bibinfo {note} {This happens when the mobility centre of a body does not coincide with its centre of mass.}\BibitemShut {Stop}%
\bibitem [{Note2()}]{Note2}%
  \BibitemOpen
  \bibinfo {note} {Note that gravity is taken to be along $-\protect \bm {\protect \hat {y}}$}\BibitemShut {NoStop}%
\end{thebibliography}%

\end{document}